\documentclass[a4paper,fleqn,usenatbib]{mnras}

\usepackage{newtxtext,newtxmath}

\usepackage[T1]{fontenc}
\usepackage{ae,aecompl}

\usepackage{graphicx}	
\usepackage{amsmath}	
\usepackage{amssymb}	

\usepackage{multicol,lipsum}
\usepackage{hyperref}
\usepackage{url}

\title[The Complex Star-forming Environment of V1647 Ori]{The ALMA Early Science View of FUor/EXor objects. IV. Misaligned Outflows in the Complex Star-forming Environment of V1647 Ori and McNeil's Nebula}

\author[D. Principe et al.]{
\hspace{-1mm}David A. Principe,$^{1,}$$^{2}$ $^{3}$\thanks{E-mail: principe@mit.edu}
Lucas Cieza, $^{2}$ $^{3}$
Antonio Hales, $^{4}$ $^{5}$
Alice Zurlo, $^{2}$ $^{3}$
\newauthor
Jonathan Williams, $^{6}$
Dary Ruiz-Rodriguez, $^{7}$
Hector Canovas, $^{8}$
Simon Casassus, $^{9}$ $^{3}$
\newauthor
Koraljka Muzic, $^{2}$ $^{10}$
Sebastian Perez,$^{9}$ $^{3}$
John J. Tobin, $^{11}$ $^{12}$
Zhaohuan Zhu $^{13}$
\\
$^{1}$Massachusetts Institute of Technology, Kavli Institute for Astrophysics, Cambridge, MA, USA\\
$^{2}$N{\'u}cleo de Astronom{\'i}a de la Facultad de Ingenier{\'i}a, Universidad Diego Portales, Av. Ej{\'e}rcito 441, Santiago 8320000, Chile\\
$^{3}$Millennium Nucleus Protoplanetary Disks, Chile\\
$^{4}$Atacama Large Millimeter/Submillimeter Array, Joint ALMA Observatory, Alonso de C\'ordova 3107, Vitacura 763-0355, Santiago - Chile \\
$^{5}$National Radio Astronomy Observatory, 520 Edgemont Road, Charlottesville, Virginia, 22903-2475, United States\\
$^{6}$Institute for Astronomy, University of Hawaii, Honolulu, Hawaii 96822\\
$^{7}$Research School of Astronomy and Astrophysics, Australian National University, Canberra, ACT 2611\\
$^{8}$ Departamento de Física Teórica, Universidad Autónoma de Madrid, Cantoblanco, 28049 Madrid, Spain\\
$^{9}$ Departamento de Astronomía, Universidad de Chile, Casilla 36-D, Santiago, Chile\\
$^{10}$ SIM/CENTRA, Faculdade de Ciencias de Universidade de Lisboa, Ed. C8, Campo Grande, P-1749-016 Lisboa, Portugal\\
$^{11}$Homer L. Dodge Department of Physics and Astronomy, University of Oklahoma, 440 W. Brooks Street, Norman, Oklahoma 73019\\
$^{12}$Leiden Observatory, Leiden University, P.O. Box 9513, 2300-RA Leiden, The Netherlands\\
$^{13}$Department of Physics and Astronomy, University of Nevada, Las Vegas, 4505 South Maryland Parkway, Las Vegas, NV 89154, USA\\
}

\date{Accepted 2017 September 03. Received 2017 September 02; in original form 2017 March 21}

\pubyear{2017}

\begin{document}
\label{firstpage}
\pagerange{\pageref{firstpage}--\pageref{lastpage}}
\maketitle

\begin{abstract}

We present Atacama Large Millimeter/sub-millimeter Array (ALMA) observations of the star-forming environment surrounding V1647 Ori, an outbursting FUor/EXor pre-MS star.  Dust continuum and the $(J=2-1)$ $^{12}$CO, $^{13}$CO, C$^{18}$O molecular emission lines were observed to characterize the V1647 Ori circumstellar disc and any large scale molecular features present.  We detect continuum emission from the circumstellar disc and determine a radius r $= 40$ au, inclination {\it i} $ = 17$$^{\circ}$$^{+6}_{-9}$ and total disc mass of M$_\mathrm{disc}$ of $\sim$0.1 M$_{\odot}$. We do not identify any disc structures associated with nearby companions, massive planets or fragmentation. The molecular cloud environment surrounding V1647 Ori is both structured and complex.  We confirm the presence of an excavated cavity north of V1647 Ori and have identified dense material at the base of the optical reflection nebula (McNeil's Nebula) that is actively shaping its surrounding environment.  Two distinct outflows have been detected with dynamical ages of $\sim$11,700 and 17,200 years.  These outflows are misaligned suggesting disc precession over $\sim$5500 years  as a result of anisotropic accretion events is responsible.  The collimated outflows exhibit velocities of $\sim$2 km s$^{-1}$, similar in velocity to that of other FUor objects presented in this series but significantly slower than previous observations and model predictions.  The V1647 Ori system is seemingly connected by an "arm" of material to a large unresolved structure located $\sim$20$''$ to the west.  The complex environment surrounding V1647 Ori suggests it is in the early stages of star formation which may relate to its classification as both an FUor and EXor type object.

\end{abstract}

\begin{keywords}
stars: evolution -- accretion -- stars:formation -- stars: mass loss -- stars:protostars -- stars: kinematics and dynamics \vspace{1in}
\end{keywords}



\section{Introduction}

\hspace {6mm} Stars form as a result of gravitational contraction of rotating molecular clouds. As the gravitational collapse proceeds, a system emerges composed of an accreting star, disc and envelope.  Observations of this stage of pre-main sequence (pre-MS) stellar evolution display large scale molecular outflows that act to transport energy back into the surrounding environment and are likely the main dispersing mechanism of molecular cloud material, responsible for the transition from a deeply embedded Class 0 protostar to the more evolved Class I/II pre-MS stellar systems.

An empirical relationship has been identified between the opening angle of these outflows and the pre-MS stage of stellar evolution \citep[Class 0/I/II;][and references therein]{Williams2011}.  The earliest stages of evolution (Class 0) exhibit outflows that are highly collimated and have opening angles of 20-50$^{\circ}$ whereas the more evolved Class I and II stages are associated with less collimated opening angles of 80-120$^{\circ}$ and 100-160$^{\circ}$, respectively \citep{Arce2006}.  The origin of these outflows are debated due to the au spatial scales required to resolve them however several theories relate the outflows to intense mass accretion events from the disc and/or envelope onto the central pre-MS star \citep[][and references therein.]{Frank2014}.  Collimated and wide-angle outflows appear to represent two distinct structures in the star-forming environment of their host stars.  In particular, wide-angle outflows tend to be more massive and slower (velocities ranging from $\sim$10-30 km s$^{-1}$) compared to highly collimated outflows which can reach velocities of $\sim$ 100-1000 km s$^{-1}$.  Their distinct velocities and masses arise from the mechanisms responsible for their production; wide angle outflows are likely produced from the ambient molecular material getting swept up when interacting with a jet bow shock whereas highly collimated jets likely arise from the interaction between a strongly magnetized central star and the inner edge of its accretion disc during intense mass accretion events \citep{Frank2014}.  Recent observations indicate that extended disc winds may also drive protostellar outflows \citep{Bjerkeli2016b}. 

Mass accretion events where material gets loaded from the disc onto the protostar are essential to star formation as these events are the primary means by which stars gain their mass.  The standard model of star formation indicates that a molecular cloud gravitationally contracts and forms a star-disc system in about $\sim$5 $\times$ 10$^{5}$ years \citep{Evans2009}.  The mass accretion rate necessary to produce a star-disc system after this timescale is inconsistent with the age and the implied accretion rates derived from the observed accretion luminosity of a large sample of stars in Taurus \citep{Kenyon1990}.  One solution to this "luminosity problem" is short-lived major accretion events of significant mass. In the past several decades, a phenomenon has been observed in a small subset of pre-MS stars that is consistent with brief and intense mass accretion events identified as "outbursts" \citep{Herbig1966,Herbig1977,Audard2014}. These outbursting pre-MS stars have been named FUor/EXor objects based on their namesakes FU Ori and EX Lup.  This phenomenon is identifiable when pre-MS stars suddenly increase in brightness by several magnitudes at optical/near-IR wavelengths as a result of intense mass accretion outbursts where accretion rates can reach as high as 10$^{-4}$ -- 10$^{-5}$ M$_{\odot}$ yr$^{-1}$ \citep{Audard2014}. FU Ori type (FUor) and EX Lup type (EXor) are generally distinguished by the frequency and length of their protostellar outbursts where FUor type objects tend to have longer (years to decades) timescale and EXors are characterized by more frequent and shorter outbursts (months-years).   

While it is evident these outbursts are caused by large mass-accretion events, it is not clear what underlying mechanism is responsible. Several theories have been proposed which include binary companions and/or massive planets generating disc instabilities \citep{Clarke1990,Lodato2004},  disc fragmentation \citep{Vorobyov2005}, and a combination of magneto rotational instability (MRI) and gravitational instabilities \citep{Armitage2001}. A detailed summary of these mechanisms and their relevant references can be found in \citet{Audard2014}.  Regardless of the underlying mechanism, it is clear that sudden intense outbursts can affect the circumstellar disc of protostars and thus, may impact planet formation \citep{Kennedy2008,Okuzumi2012}.  Recently, an analysis of high resolution ALMA observations of the outbursting FUor type object V883 Ori revealed the first detection of a water snow line in a circumstellar disc \citep{Cieza2016}.  Such a discovery was feasible because the water snow line, which typically resides < 5 au from a $\sim$1 M$_{\odot}$ central star, was temporarily extended to $\sim$40 au in response to the increase in luminosity of the FUor object during outburst.

V1647 Ori, a protostar in the Lynds 1630 (L1630) region of the Orion Molecular Cloud \citep[][and references therein]{Principe2014}, was first noted in 2003 when a large ($\sim$1$'$) reflection nebula (McNeil's Nebula) appeared coincident with a sudden increase in optical stellar brightness \citep[$\Delta$I $\sim$ 5 magnitudes;][]{McNeil2004, Reipurth2004, Briceno2004}.  After its identification in 2003, two more outbursts were identified: a 1966 outburst identified using archival photometric plates \citep{Aspin2006} and a 2008 outburst which increased the flux values of V1647 Ori to similar values as in 2003. During the 2008 outburst, \citet{Ninan2013} carried out a 4.5 year photometric study of V1647 Ori and report that at the end of their campaign in 2012, V1647 Ori was still in outburst after almost half a decade.  The frequency and length of these outbursts, combined with several spectral features characteristic of both FUors and EXors emphasize the ambiguity between the FUor or EXor classification of V1647 Ori \citep[][and ref. therein]{Audard2014}.  Spectral parameters of V1647 Ori were determined from data taken during quiescence and presented in \citet{Aspin2008}. They identify V1647 Ori as a young (< 0.5 Myr) pre-MS star of spectral type M0 $\pm$ 2 with an approximate bolometric luminosity, mass and radius of  $\sim$ 5.2 L$_{\odot}$, 0.8 M$_{\odot}$, and 5 R$_{\odot}$, respectively.  During its 2003 outburst, its bolometric luminosity was measured to be L$_\mathrm{bol}$ = 44 L$_{\odot}$ \citep{Muzerolle2005}.

Given the transient nature of V1647 Ori, several multiwavelength campaigns were setup to observe the star in its outbursting state. The 2003 outburst was characterized by a $\sim$3 magnitude increase in the near-IR \citep{Reipurth2004}, a factor of 25 increase in 12 $\mu$m flux \citep{Andrews2004} and a factor of 50 increase in X-ray flux \citep{Kastner2004b}.  \citet{Andrews2004} report no apparent changes in the (350 $\mu$m-1.3mm) submm dust continuum brightness during outburst compared with its quiescent state.  Plasma temperatures derived from the X-ray imaging spectrum presented in \citet{Kastner2004b} were too high to be produced via accretion alone suggesting extreme changes in magnetic field configuration (i.e., magnetic reconnection events) were be present.  Spectroscopic measurements during the 2003 outburst indicated accretion rates of a few 10$^{-6}$ to 10$^{-5}$ M$_{\odot}$ year $^{-1}$ \citep{Muzerolle2005} and H$\alpha$ P Cygni profiles indicate wind velocities up to $\sim$600 km s$^{-1}$ \citep{Reipurth2004}. Mid-IR interferometric observations probing the circumstellar disc indicate a moderately flaring disc with no signatures of close companions at radii < 100 au \citep{Abraham2006}. 

All three outbursts re-illuminated the reflection nebula, where dust grains preferentially scatter blue light, just north of V1647 Ori and several studies have identified small variations in the local environment as probed by this extended nebular emission \citep{Aspin2006,Ninan2013}.  The time delay between brightness variations of the star and of clumps in the nebulosity were used to estimate a stellar inclination angle of $\sim$61$^{\circ}$ \citep[the inclination of the polar axis towards our line of sight][]{Acosta-Pulido2007}. An X-ray periodicity study of V1647 Ori found a similar stellar inclination angle of {\it i} $\sim$68$^{\circ}$ based on modeling the rotation of accretion hot spots on the stellar surface \citep{Hamaguchi2012}.  These inclinations are in disagreement with studies that suggest a more face-on inclination of $<$ 30$^{\circ}$ based on the absence of a 9.7 $\mu$m amorphous silicate feature \citep{Andrews2004} and a low column density of CO in absorption from the disc \citep{Rettig2005}. 

 Here, we present ALMA 1.3 mm observations of the gas and dust surrounding V1647 Ori in order to characterize its circumstellar disc and local star-forming environment.  V1647 Ori is one of 8 FUor/EXor type objects observed as part of an ALMA Cycle 2 project (2013.1.00710.S ; PI L. Cieza) to investigate star-forming environments at this critical stage of pre-MS stellar evolution.  All other FUor/EXor objects observed as part of this program, including V883 Ori, HBC 494, V2775 Ori, NY Ori, V1143 Ori, V1118 Ori, ASASSN-13db, are presented in papers I, II, III, IV (this paper) and V \citep{Zurlo2017,Ruiz-Rodriguez2017a,Ruiz-Rodriguez2017b,Cieza2017}.  These recently published results indicate that the star-forming environments of these objects are structured and complex. In particular, features such as rings/shells and both wide-angle and collimated outflows have all been identified. Outflow velocities measured for V883 Ori, HBC 494, V2775 Ori, and V1647 Ori (Section 3.6) as part of this series of papers indicate that significantly slower velocities (v$\sim$ 1 - 4 km $s^{-1}$) are present than those typically observed and associated with protostellar outflows at these wavelengths \citep[][ and references therein]{Frank2014}.  The resolving power and sensitivity of the observations presented in this series also provides strict constraints on circumstellar disc characteristics such as disc size, orientation, and mass.

\section{Observations} 

V1647 Ori was observed with ALMA band-6 at three epochs  (December 12th, 2014, April 5th, 2015 and August 30th, 2015).  The first two observations were performed with an ALMA configuration of 45 antennas (12 meter diameter) and baselines ranging from 14.6 to 348.5 meters which achieved an angular resolution of $\sim$1.0 $''$.  The final epoch had a configuration of 35 antennas and longer baselines of 42-1574 meters resulting in a higher angular resolution of 0.2 $''$. The precipitable water vapor (PWV) levels were 0.7, 1.3 and 1 mm for the December 2014, April 2015, and August 2015 observations, respectively.  Each ALMA observation of V1647 Ori include two broad (2 GHz) continuum  spectral windows centered at $\sim$232 GHz and $\sim$218 GHz and three narrow (59 MHz) windows centered near the rest frequencies of $^{12}$CO (J = 2-1; 230.5380 GHz), $^{13}$CO (J = 2-1; 220.3987 GHz) and C180 (J = 2-1; 219.5603 GHz). Flux calibration was performed with Ganymede and J0423-013 while bandpass calibration was performed with quasars J0538-4405 and J0607-0834.  The time dependence variations of the complex gains were calibrated by alternating V1647 Ori with nearby phase calibrators J0541-0541, J0532-0307 and/or J0529-0519.

The ALMA pipeline-calibrated observations were further reduced with the Common Astronomical Software Application \citep[CASA v4.7;][]{McMullin2007}.  The two epochs of $\sim$1.0$''$ resolution data were continuum subtracted in the visibility domain, concatenated, and cleaned to form a single dataset for the continuum and each spectral line.  The same procedure was performed for the single epoch of $\sim$0.2$''$ resolution data. While features of the V1647 Ori disc/outflows are clearly present in the $\sim$1$''$ resolution datasets, none of these features are identifiable in the high resolution line imaging from the 0.2$''$ resolution dataset.  This is consistent with most of the flux from the system being fairly uniform on large scales. Therefore we only present the detection of continuum emission from the 0.2$''$ dataset and the line and continuum emission from the combined two $\sim$1$''$ datasets.  The beam size, position angle, and 3$\sigma$ sensitivities for each spectral line channel map and the continuum are indicated in Table \ref{observation_table}. 

Integrated intensity images (moment 0) were creating using the CASA routine $immoments$. The moment 0 images were created using only pixels with a signal detection of 3$\sigma$ or higher.  To better describe the morphology and dynamics of the V1647 Ori star-forming environment, we created moment 0 images for each spectral line with three distinct velocity ranges: $5.0-9.49$ km s$^{-1}$, $9.5-10.5$ km s$^{-1}$ and $10.51-13.0$ km s$^{-1}$.  These velocity bands will henceforth be referred to in this paper as blueshifted, systemic, and redshifted, respectively. The images presented in this work are not primary-beam corrected. However, primary-beam corrected channel maps were used when estimating physical parameters of the outflow from the line fluxes (Section 3.6).
 
Optical Gemini-GMOS imaging observations of V1647 Ori were performed on September 22, 2008 with an exposure time of 60 seconds in R band. Optical imaging data was retrieved from the Gemini Science Archive and presented here to supplement the ALMA observations.  Optical analysis of these and similar data are presented in other work \citep{Aspin2009b} and the discussion in this paper is limited only to correlations between large-scale optical and mm emission morphology.

\begin{table*}
\centering
\caption{V1647 Ori ALMA Continuum and Integrated Intensity Image Parameters}

{\renewcommand{\arraystretch}{1.0}
{
\begin{tabular}{ c | c l c | c l }

\hline
	 & Median Restoring Beam & Position Angle & Sensitivity (3$\sigma$) \\ 
	 & [$''$] & \hspace{7mm}[$^{\circ}$] &  \\ \hline
	Continuum & 0.26$\times$0.20  &\hspace{3.5mm} \hspace{0.9mm}88.0   & 0.80 [mJy beam$^{-1}$]  \\
	$^{12}$CO & 1.61$\times$0.98  &\hspace{3.5mm} -87.21 & 45 [mJy km s$^{-1}$ beam$^{-1}$]  \\ 
	$^{13}$CO & 1.66$\times$1.03  &\hspace{3.5mm} -85.58  & 45 [mJy km s$^{-1}$ beam$^{-1}$]\\ 
	C$^{18}$O & 1.69$\times$1.03  &\hspace{3.5mm} \hspace{0.9mm}87.42  & 36 [mJy km s$^{-1}$ beam$^{-1}$]  \\ \hline
\end{tabular}
}}
\label{observation_table}
\end{table*}

\section{Results}

\subsection{Continuum Emission}

\begin{figure}
\centering
\includegraphics[scale=0.45]{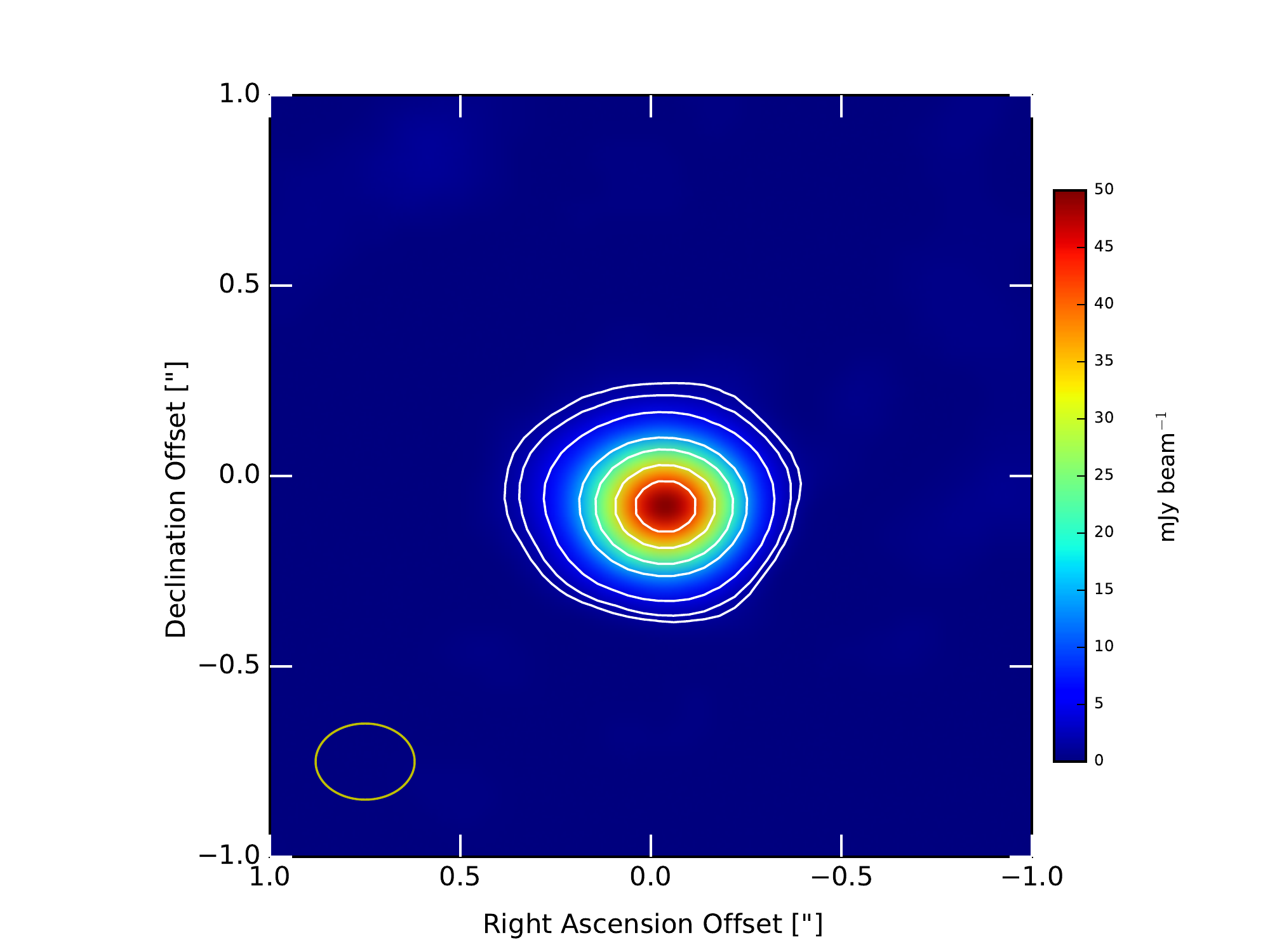}
\caption{Resolved continuum emission of the V1467 Ori circumstellar disc with contours overlaid of values 1.75, 3, 7, 20, 30, 45 and 60 $\times$ $3\sigma$. A  two-dimensional gaussian fit indicates a disc with inclination of 17$^{\circ}$$^{+6}_{-9}$ and PA = 109$\pm19$$^{\circ}$.  The beam size is displayed in the bottom left and north is up and east is left.}
\label{continuum}
\end{figure}

The integrated intensity image of the resolved 0.2$''$ V1647 Ori continuum is displayed in Figure \ref{continuum}. Two-dimensional gaussian fitting of the $\sim$225 GHz continuum emission performed in CASA indicate a flux, inclination, position angle and centroid of 82.62 $\pm$ 8.26 mJy, ${\it i} = 17^{\circ}$ $^{+6}_{-9}$, PA$ = 109^{\circ} \pm$ $19$, and RA= 05:46:13.139, Dec= -00:06:04.90, respectively. The inclination was calculated using the aspect ratio of an assumed circular disc. All position angles reported in this work are east of north.  A semi major axis FWHM angular size of 189.0 $\pm$ 3.6 milliarcseconds and a distance of 414 $\pm{7}$ pc \citep{Menten2007} results in a disc dust radius of $\sim$40 au. Evidence that the continuum emission from the circumstellar disc is spatially resolved is identified by its amplitude as a function of UV distance and asymmetry in the continuum emission (Figure \ref{continuum}) indicates a potentially non-gaussian distribution near the outer edges of the image.   A more in-depth analysis of continuum emission from V1647 Ori as well as the other FUor/EXor objects in this series will be presented in \citet{Cieza2017}.

The mass of the V1647 Ori circumstellar disc dust can be estimated from the continuum emission (assuming continuum emission is thermal and optically thin) following \citet{Beckwith1990} where:
\begin{equation}
\hspace{1.2in}M_{\rm{disc}}= \frac{F_\nu d^2}{\kappa_\nu B_\nu(T)},
\end{equation}
with $B_\nu(T)$ representing the Planck function for dust at a temperature $T$ and $\kappa_\nu$ as the dust opacity, which is a power-law function of submm frequency, i.e., $\kappa_\nu=0.1(\nu/ 1000$ $\mathrm{GHz})^\beta$ cm$^{2}$ g$^{-1}$.  The power law opacity index $\beta$  is related to the size distribution and composition of the disc dust particles. While $\beta$ can range from $\beta$=0 for certain grain mineralogies or size distributions \citep{Pollack1994} to $\beta$=2 for interstellar dust grains in the ISM \citep{Schwartz1982}, it is more likely to be in the range of 0.5-1.5 for circumstellar discs \citep{Andrews2005}. Without more sophisticated continuum modeling to estimate the disc dust temperature of V1647 Ori, a dust temperature and range of $T=20 \pm 10$ K is chosen based on the median temperature of discs in Taurus-Auriga \citep{Andrews2005} and the relatively small temperature range indicated in modeling of other circumstellar discs \citep{Tazzari2017}. Assuming values of $\beta$=1, d= 414 pc, $\kappa_\nu$ = 0.023 cm$^{2}$ g$^{-1}$, and a dust temperature of $T=20$ K, we estimate the disc dust mass of V1647 Ori to be $M_\mathrm{dust}$ = 427 M$_\mathrm{E}$. This corresponds to a total disc mass $M_\mathrm{disc}$ = 0.13 $M_{\odot}$ ($\sim$136 M$_\mathrm{Jup}$) assuming a gas to dust mass ratio of 100. When converting from flux to disc mass using Equation 1, the largest sources of uncertainty are the ranges in $\beta$ and dust temperature. While the uncertainty in the continuum flux is dominated by the 10\% absolute flux uncertainty of ALMA Band 6, the potential range in $\beta$ ($\beta$= 1 $\pm{0.5}$) and dust temperature (T = 20 $\pm{10}$) introduce much higher uncertainties in dust mass that can result in masses in the extreme ranges of M$_{\rm{dust}}$ = [110 - 2688] M$_\mathrm{E}$ or $M_{\rm{disc}}$ = [0.03 - 0.80] $M_{\odot}$ assuming a gas to dust ratio of 100.  A summary of V1647 Ori disc parameters are displayed in Table \ref{continuum_table}.

\begin{table}
\centering
\caption{V1647 Ori Disc Parameters Derived From Continuum Emission}


{\renewcommand{\arraystretch}{1.0}

{
\begin{tabular}{ c  c   }
\hline

 & V1647 Ori  \\ \hline
Right Ascension & 05:46:13.14  \\
Declination & -00.06.04.90  \\
Flux [mJy] & 82.62 $\pm$ $8.26$  \\ 
Inclination & 17$^{\circ}$ $^{+6}_{-9}$  \\
Position Angle & 109$^{\circ}$ $\pm $ $19$  \\ 
Disk Radius [au] & 40    \\
Disc Dust Mass [M$_{E}$] & 427 $\pm$ \footnote{1}$43$ \\
Disc Total Mass\footnote{2} [M$_{\odot}$] & 0.13 \\ \hline
\end{tabular}}}
\hspace{3in}\footnotemark[1]{This uncertainty is based only on the 10\% absolute flux uncertainty of ALMA Band 6 and may be up to a factor of five higher than the dust mass if considering the uncertainty introduced by the range in disc temperature and opacity index (see Section 3.1). }
\hspace{3in}\footnotemark[2]{Estimated assuming a gas to dust mass ratio of 100.}
\label{continuum_table}
\end{table}

\subsection{Gas Emission Lines}

\begin{figure}
\centering
\includegraphics[scale=0.2]{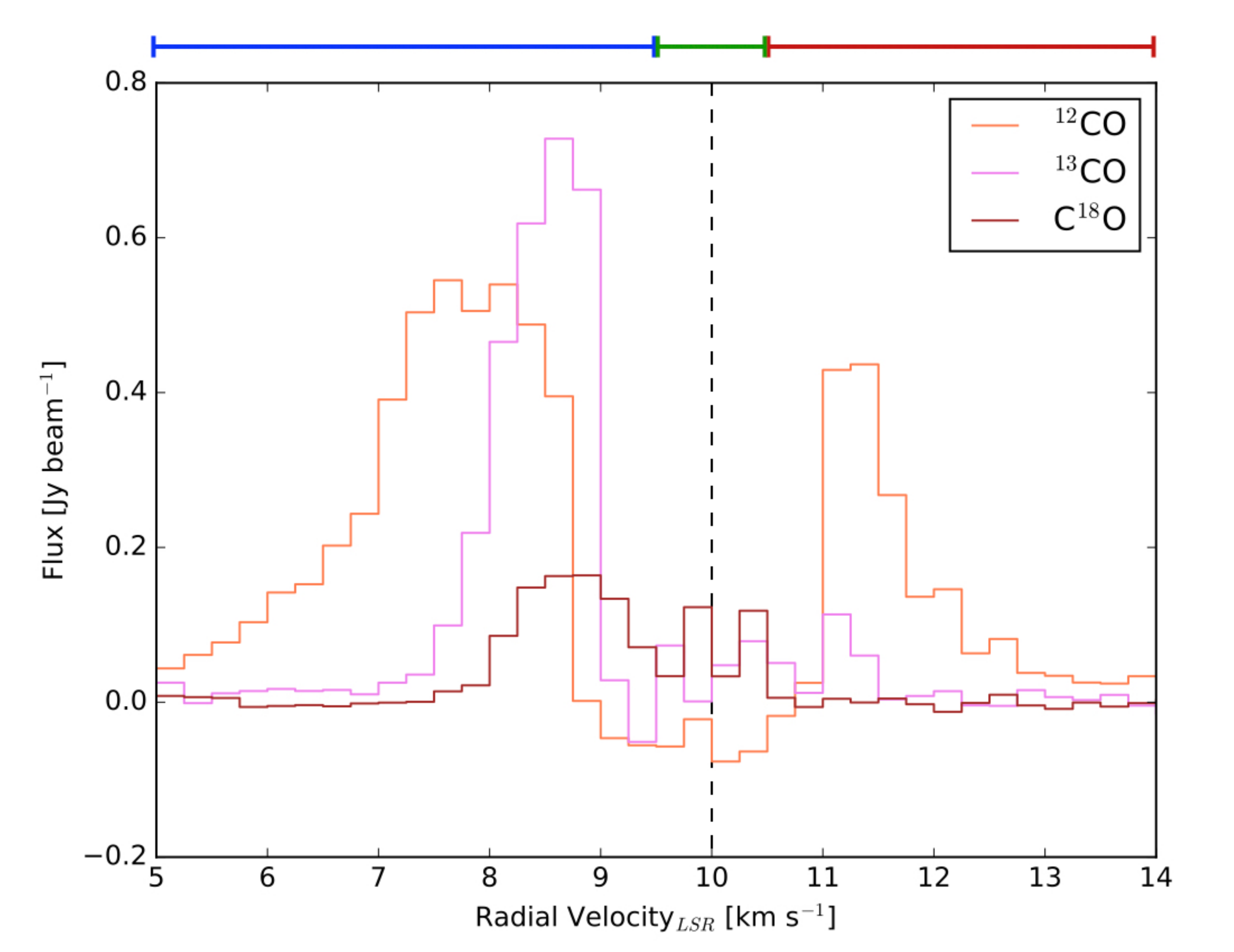}
\includegraphics[scale=0.2]{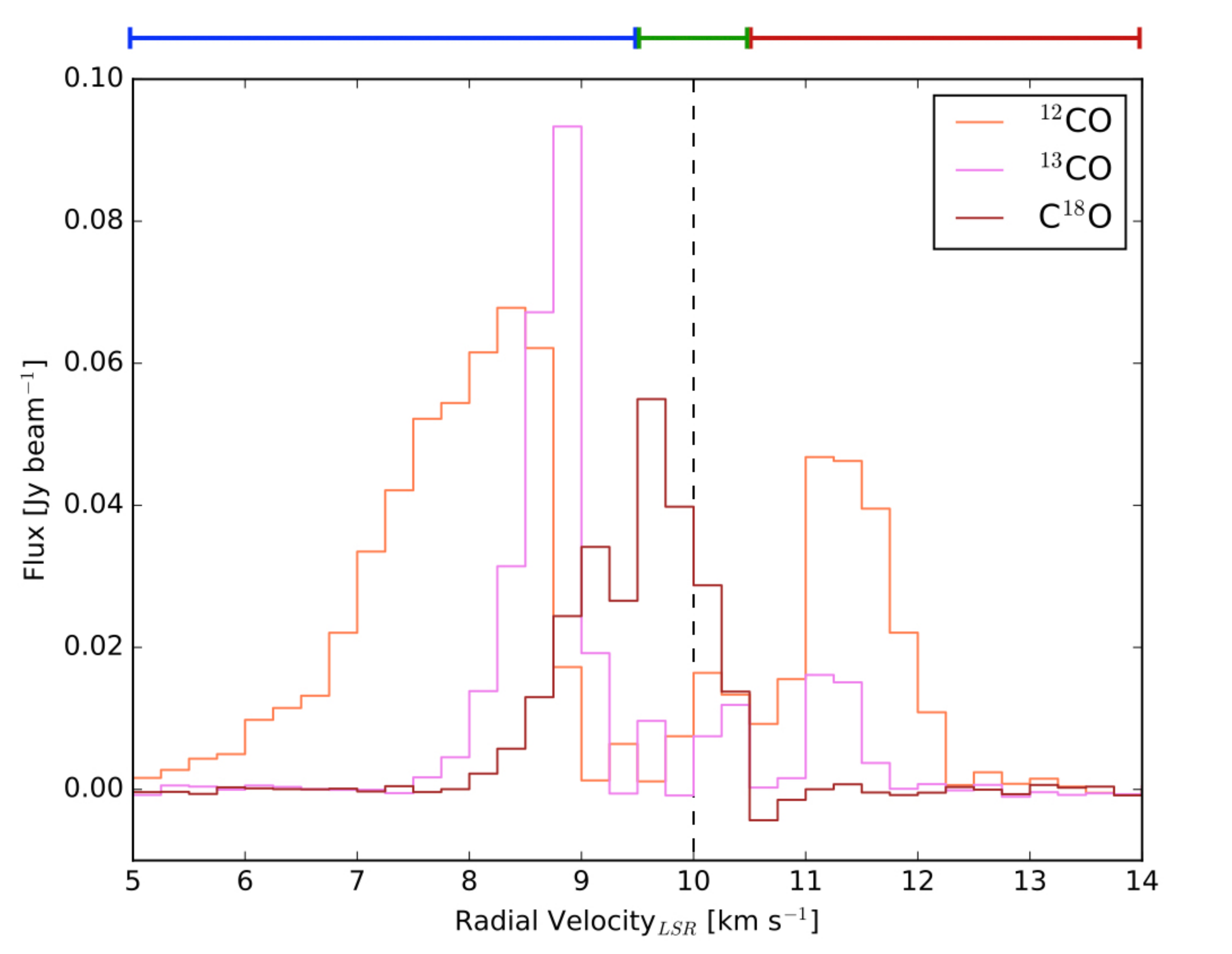}
\caption{ Line profiles from apertures centered on the dust continuum emission of V1647 Ori with diameters of 2$''$ (top) and 20$''$ (bottom) and velocity bins of 0.25 km s$^{-1}$. A systemic velocity of 10.0 km s$^{-1}$ is indicated with a vertical dashed line. The blue, green and red lines above the plot represent the velocity ranges described in Section 2 for the blueshifted, systemic, and redshifted images presented throughout the rest of this paper.  Individual line channel maps are included in the Appendix.}
\label{spec_lines}
\end{figure}

\begin{figure*}
\centering
\includegraphics[scale=0.85]{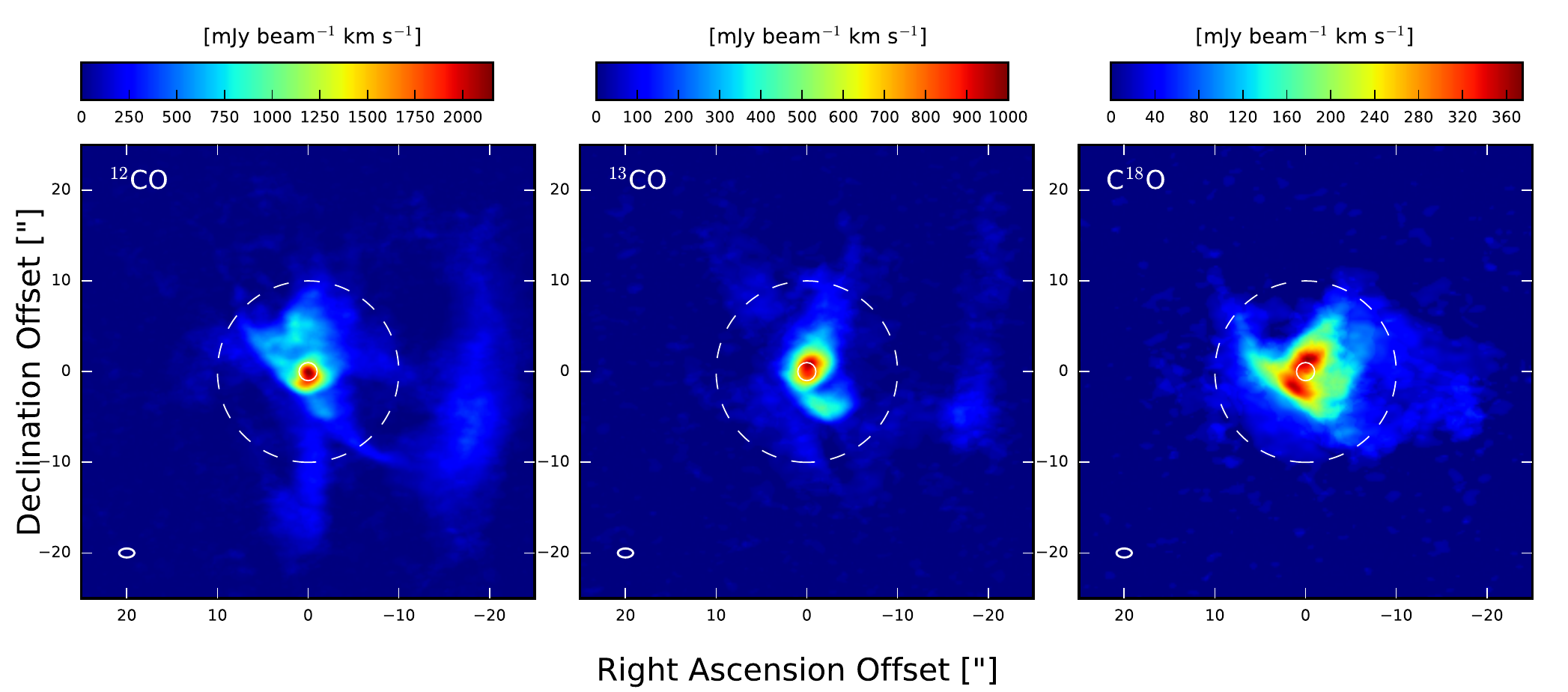}
\caption{ Integrated intensity (moment 0) images of $^{12}$CO, $^{13}$CO and C$^{18}$O integrated over each lines entire velocity range.  V1647 Ori is at the center of each panel (offset [0,0] corresponds to RA= 05:46:13.135, Dec= -00:06:04.82).  The small solid and large dashed white lines indicate the spectral extraction regions from Figure \ref{spec_lines} (aperture diameters of 2$''$ and 20$''$, respectively) with north is up and east is left. The beam is displayed in the lower left.}
\label{mom0_all_lines}
\end{figure*}

\begin{figure}
\centering
\includegraphics[scale=0.35]{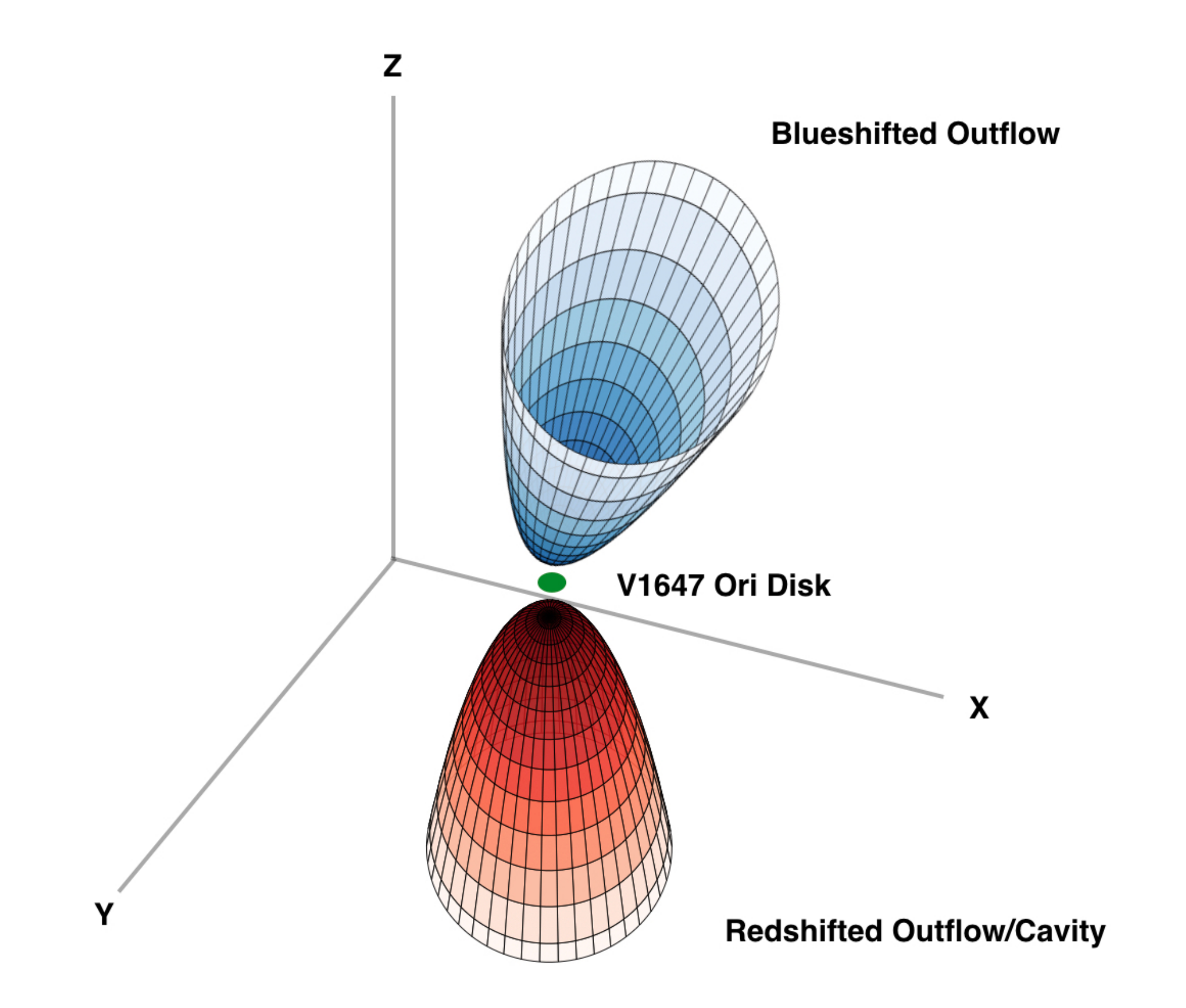}
\caption{ A three-dimensional illustration of V1647 Ori and its surrounding misaligned outflows (labels D and E discussed in Sections 3.3 and 3.4). The nearly face-on circumstellar disc is shown in green.  Features in the illustration are not to scale. }
\label{3d}
\end{figure}

Spectral flux density as a function of radial velocity for the $^{12}$CO, $^{13}$CO, and C$^{18}$O lines are displayed in 0.25 km s$^{-1}$ bins in Figure \ref{spec_lines}. Spectral extraction was performed using two different apertures both centered at the position of V1647 Ori (J2000 RA= 05:46:13.135, Dec= -00:06:04.82).  An aperture diameter of 2.0 $''$ was chosen to extract spectral information at the precise location of the continuum emission (Fig \ref{spec_lines} top) and an aperture diameter of 20$''$ was chosen to encompass the nebulosity surrounding V1647 Ori (Fig \ref{spec_lines} bottom).  In both the small and large apertures, a double peaked $^{12}$CO and single peaked $^{13}$CO emission feature is observed.  A clear detection of C$^{18}$O material is seen in the large 20$''$ aperture indicating an abundance of dense material traveling at a radial velocity equidistant between the two $^{12}$CO peaks.  Due to the embedded nature of this object, the rest velocity of the system (i.e., systemic velocity) is not clear.  In the absence of a literature value of the systemic velocity of V1647 Ori, we estimate a systemic velocity of $\sim$10.0 km s$^{-1}$ assuming the star forms in the densest part of the molecular cloud (i.e. at the same velocity as the bulk of the dense C$^{18}$O emission; Figure \ref{spec_lines}).  This systemic velocity is also consistent with being equidistant between the peaks of the blue and red shifted $^{12}$CO emission lines. The blueshifted, systemic, and redshifted velocity ranges are indicated at the top of Figure \ref{spec_lines}.  Individual channel maps for each spectral line where emission associated with the V1647 Ori environment is observed is shown in Appendix Figures 1-3.  The integrated intensity images of $^{12}$CO, $^{13}$CO, and C$^{18}$O emission over their entire velocity range are displayed in Figure \ref{mom0_all_lines} with white circles corresponding to the extraction apertures associated with the line profile displayed in Figure \ref{spec_lines}.  Several features are readily identifiable and will be discussed in detail in the following sections. An illustration of the three-dimensional system geometry is displayed in Figure \ref{3d}.

\subsection{Emission from $^{12}$CO}

The $^{12}$CO emission line is typically optically thick in molecular cloud environments and is generally not a good probe of column density.  Instead, $^{12}$CO emission traces molecular material of a particular temperature and is useful for probing excavated cavities as well as wide-angle outflows \citep{Zurlo2017,Ruiz-Rodriguez2017a}. Figure \ref{$^{12}$CO_mom0_shifts} displays the blue, systemic and red shifted $^{12}$CO emission summed over the velocity ranges indicated in the top of Figure \ref{spec_lines} and in Section 2.  The base of McNeil's Nebula, the familiar reflection nebula first identified from optical observations during its 2003 outburst is clearly seen in the blue shifted $^{12}$CO emission line (label A).  This northern cavity has a well-defined wall and an opening angle to the north of $\sim$100$^{\circ}$ which extends $\sim$ 10$''$ ($\sim$4140 au).  A bright extended region spatially coincident with the star-disc system is identifiable at radii of $r \lesssim 3''$ (1245 au) from V1647 Ori.  A distinct clump of material is seen $\sim$2$''$ south-west from the central emission (label B) appearing to be cut off from the main nebula.  Meanwhile, the redshifted morphology of the $^{12}$CO emission line is distinct from the blueshifted emission.  Emission coincident with the position of the circumstellar disc of V1647 Ori is detected in the arc-like shape bending from the north-west to the south-west.  Such a well defined structure is seemingly connected with an 'arm' to an atypically large column of $^{12}$CO emission located 20$''$ (8300 au) west of V1647 Ori near the edge of the ALMA field of view (label C).   The northern bowl-like nebula may be faintly detected behind the arc-like feature in the red-shifted $^{12}$CO emission.  More prominently displayed is an apparent southern feature extending in a straight column south of the V1647 Ori disc with an angular length of $\sim$20$''$ ($\sim$8300 au; label D ). Parts of both the large southern and western columns are identified in the blueshifted $^{12}$CO emission (Figure \ref{$^{12}$CO_mom0_shifts}; label D).  The systemic $^{12}$CO emission of V1647 Ori appears to be resolved out or undetected and instead displays features that may originate from foreground cloud contamination.

\begin{figure*}
\centering
\includegraphics[scale=0.28]{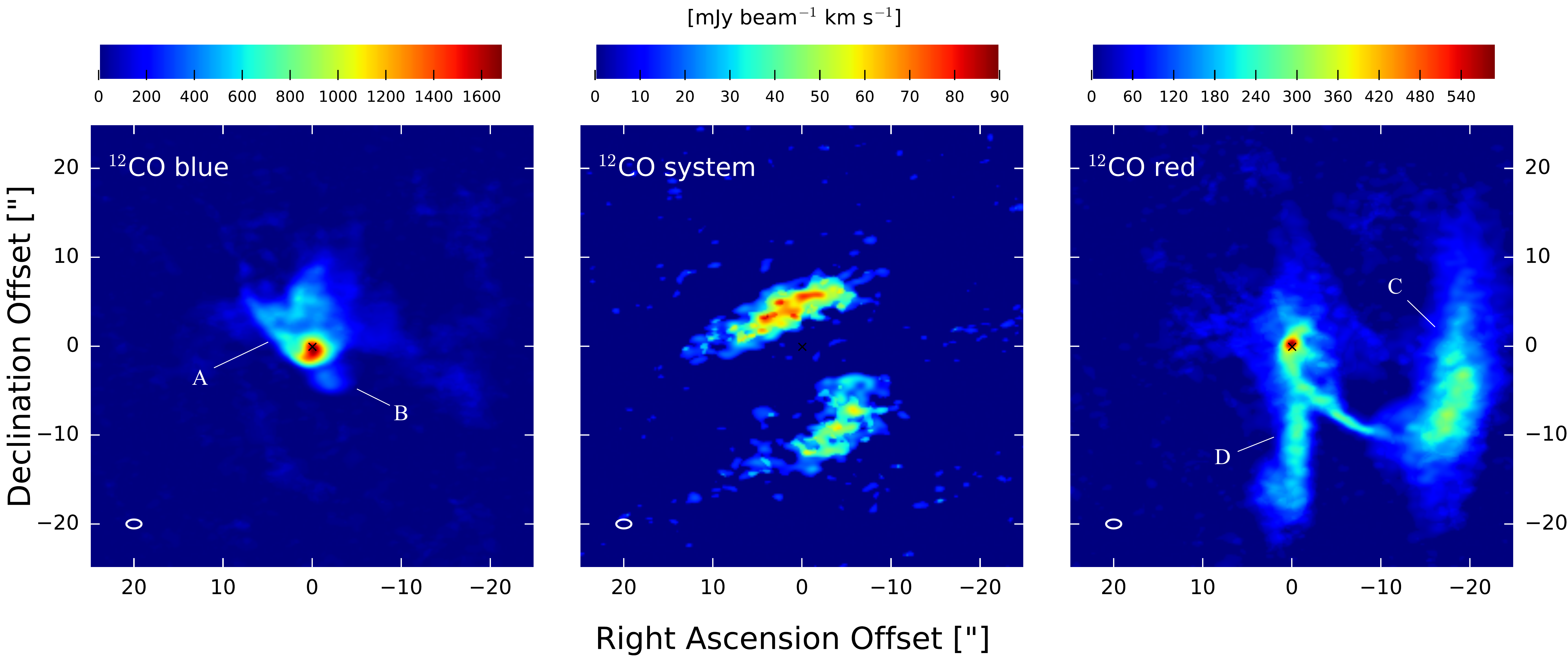}
\caption{ Side by side comparison of blue, systemic, and red shifted moment 0 images of $^{12}$CO. The central location of the V1647 Ori disc continuum (Fig \ref{continuum}) is shown with a black cross and the beam is displayed in the lower left. Figures are labeled according to the text in Section 3.  All $^{12}$CO channels associated with emission can be seen in Figure 1 of the Appendix.}
\label{$^{12}$CO_mom0_shifts}
\end{figure*}

\subsection{Emission from $^{13}$CO}

Unlike $^{12}$CO, typical densities of $^{13}$CO in molecular cloud environments are low enough such that the emission line is optically thin and is a better indicator of column density.   As such, the blue and redshifted emission features displayed in Figure \ref{$^{13}$CO_mom0_shifts} likely trace material outflowing away from the central star whereas the systemic emission features likely represent material located near the central star-disc system, out of which the star has been forming.  Similar to the case of the $^{12}$CO blueshifted emission, a clump of material is identifiable in the blueshifted $^{13}$CO emission at position angle east of north of $\sim$220$^{\circ}$ seemingly disconnected or 'pinched' from the central region (label B). The prominent blueshifted outflow feature (Figure \ref{$^{13}$CO_mom0_shifts} left) identifiable at position angle $\sim$330$^{\circ}$ (label E) has an opening angle of $\sim$50$^{\circ}$ extending to distances of $\sim$11$''$ ($\sim$4500 au). The peak intensity in the blueshifted image is located $\sim$1$''$ north-west of V1647 Ori, in the direction of the outflow.   A second collimated outflow is detected in the redshifted line emission pointing almost directly south and extending a distance of 12.8$''$ ($\sim$5300 au; label D).  Similar to the case of the blueshifted emission, the peak intensity of the redshifted image is located $\sim$1$''$ south-east of V1647 Ori, in the direction of the southern outflow.  The systemic $^{13}$CO emission feature (Figure \ref{$^{13}$CO_mom0_shifts} center) displays an apparent "hole" coincident with the location of the collimated $^{13}$CO blueshifted outflow (label F).

\begin{figure*}
\centering
\includegraphics[scale=0.28]{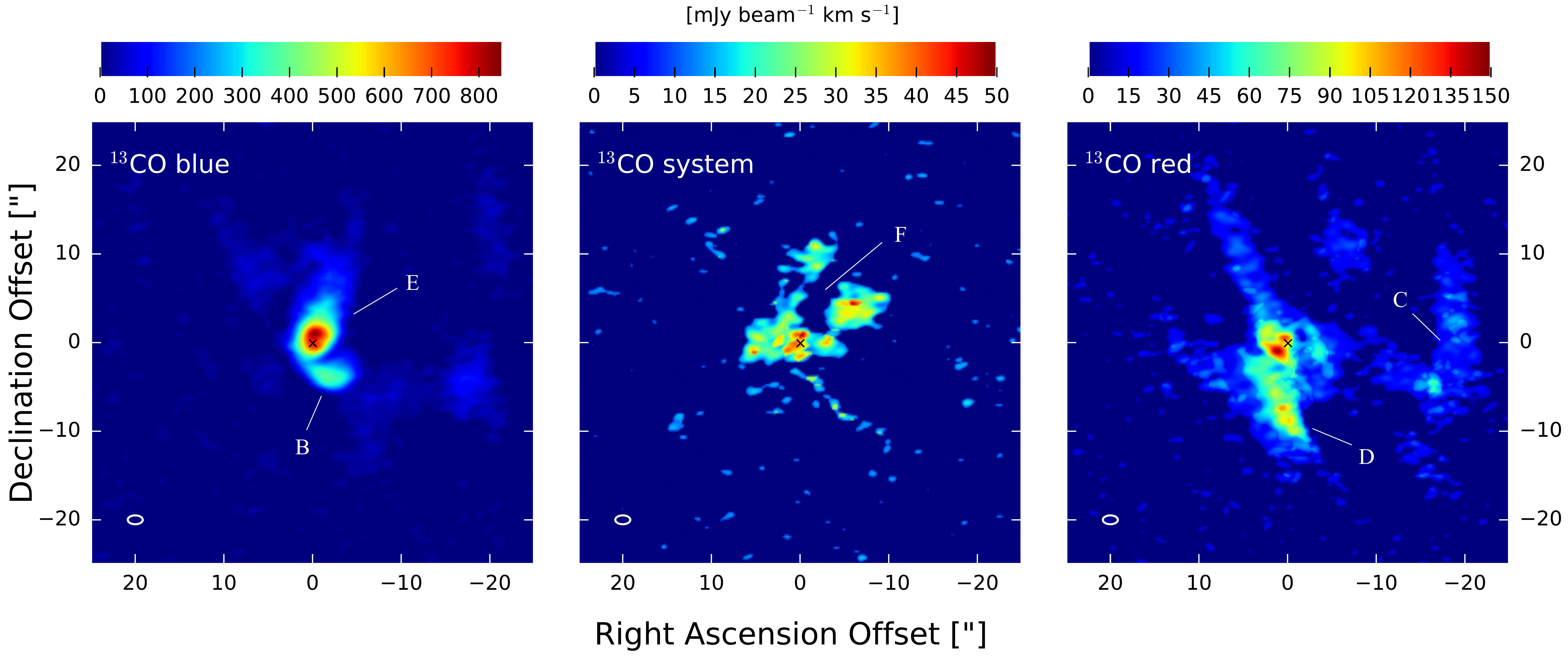}
\caption{ Side by side comparison of blue, systemic, and red shifted moment 0 images of $^{13}$CO. The central location of the V1647 Ori disc continuum (Fig \ref{continuum}) is shown with a black cross. Figures are labeled according to the text in Section 3. All $^{13}$CO channels associated with emission can be seen in Figure 1 of the Appendix. }
\label{$^{13}$CO_mom0_shifts}
\end{figure*}

\subsection{Emission from C$^{18}$O}

 C$^{18}$O emission, typically a tracer of higher column density gas in molecular cloud environments, is displayed in Figure \ref{C$^{18}$O_mom0_shifts}.  Distinct features are seen in the blue, systemic and red shifted velocity channels.  The blue shifted C$^{18}$O features trace the collimated outflowing material (label E) previously identified in the $^{13}$CO integrated intensity images (Figure \ref{$^{13}$CO_mom0_shifts}). This feature has roughly the same opening angle and angular size as its blueshifted $^{13}$CO counterpart.  Similar to the case of the blueshifted $^{13}$CO emission, the peak intensity location in the blueshifted C$^{18}$O is located $\sim$1$''$ northwest of V1647 Ori in the direction of the blueshifted outflow. The same clump located at position angle $\sim$220 from the $^{12}$CO and $^{13}$CO blueshifted integrated intensity images, apparently disconnected ('pinched') from the main emission features of V1647 Ori, is also observed to the south west in blueshifted C$^{18}$O emission (label B).  While the redshifted C$^{18}$O emission does not display the same collimated feature as is evident in the $^{13}$CO redshifted emission, it does however display two bright peaks, one of which is coincident with the location of the peak intensity of the $^{13}$CO redshifted emission south east of V1647 Ori (i.e., in the direction of the southern collimated outflow).  The second bright peak in the C$^{18}$O redshifted image is very close to the location of the V1647 Ori disc and may be a detection of circumstellar disc gas.

\begin{figure*}
\centering
\includegraphics[scale=0.28]{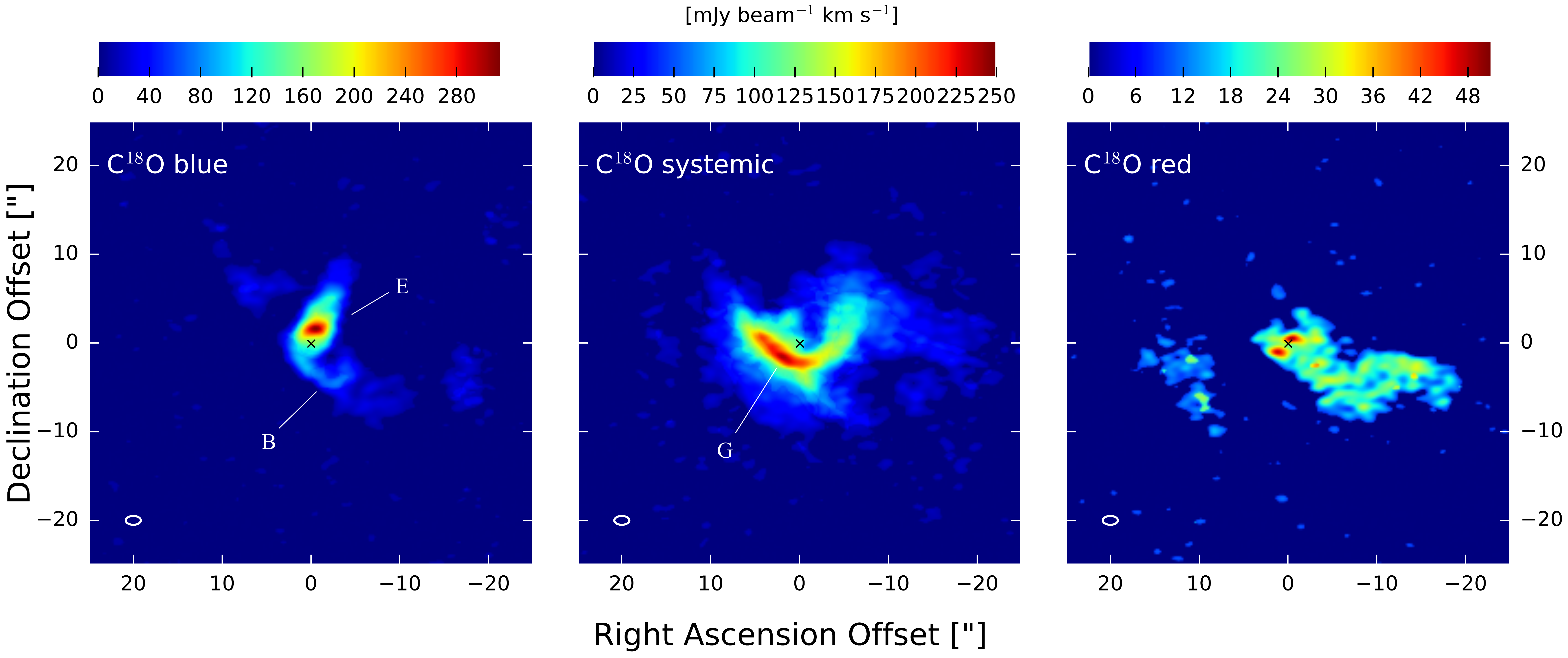}
\caption{ Side by side comparison of blue, systemic, and red shifted moment 0 images of C$^{18}$O. The central location of the V1647 Ori disc continuum (Fig \ref{continuum}) is shown with a black cross. Figures are labeled according to the text in Section 3.  All C$^{18}$O channels associated with emission can be seen in Figure 1 of the Appendix.}
\label{C$^{18}$O_mom0_shifts}
\end{figure*}

A strong detection of C$^{18}$O at systemic velocities likely represents the dense material from which the V1647 Ori protostar formed and may impact the structures identifiable in the other spectral lines.  To emphasize this role, the systemic C$^{18}$O emission contours are overlaid on blueshifted $^{12}$CO, $^{13}$CO, and C$^{18}$O integrated intensity images in Figure \ref{C$^{18}$O_system_contours}.  The systemic C$^{18}$O emission rests as the base of the bowl identified in blueshifted $^{12}$CO (Figure \ref{C$^{18}$O_system_contours} left) and also is spatially coincident with an apparent discontinuity between the blueshifted $^{12}$CO, $^{13}$CO, and C$^{18}$O emission at position angle $\sim$220$^{\circ}$ (label B; i.e., these emission features appear to 'bend' around the material probed by systemic C$^{18}$O emission).  Moreover, the large ($\sim$10$''$) bright extended emission feature in the south east part of the systemic C$^{18}$O emission appears to be in line with the axis of the $^{13}$CO and C$^{18}$O collimated outflow (label G in Figure \ref{C$^{18}$O_mom0_shifts}).  The redshifted C$^{18}$O material may also play a role in shaping of the nebula as displayed in Figure \ref{contours_extended_v2}. Redshifted C$^{18}$O emission material may be shaping the long extended arm observed in redshifted $^{12}$CO that is  apparently connecting the V1647 Ori system to an extended feature at a projected distance of $\sim$8300 au.

\begin{figure*}
\centering
\includegraphics[scale=0.75]{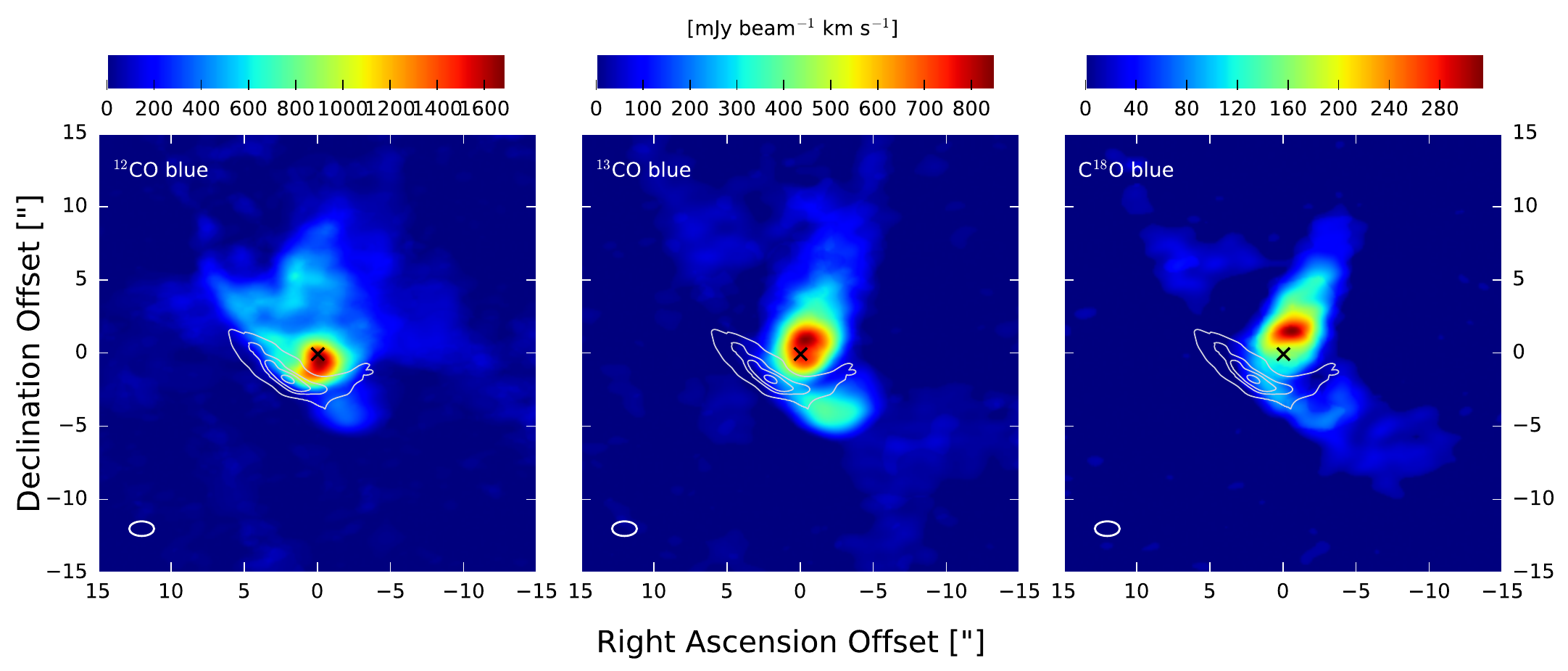}
\caption{C$^{18}$O systemic emission (white contours) is shown overlaid on blueshifted $^{12}$CO, $^{13}$CO and C$^{18}$O integrated intensity images.  The contour levels for the systemic C$^{18}$O emission are 4.0, 5.5, 6.0, 6.5, and 6.75 $\times$ 3$\sigma$.  The beam is displayed in the lower left. }
\label{C$^{18}$O_system_contours}
\end{figure*}

\begin{figure}
\centering
\includegraphics[scale=0.4]{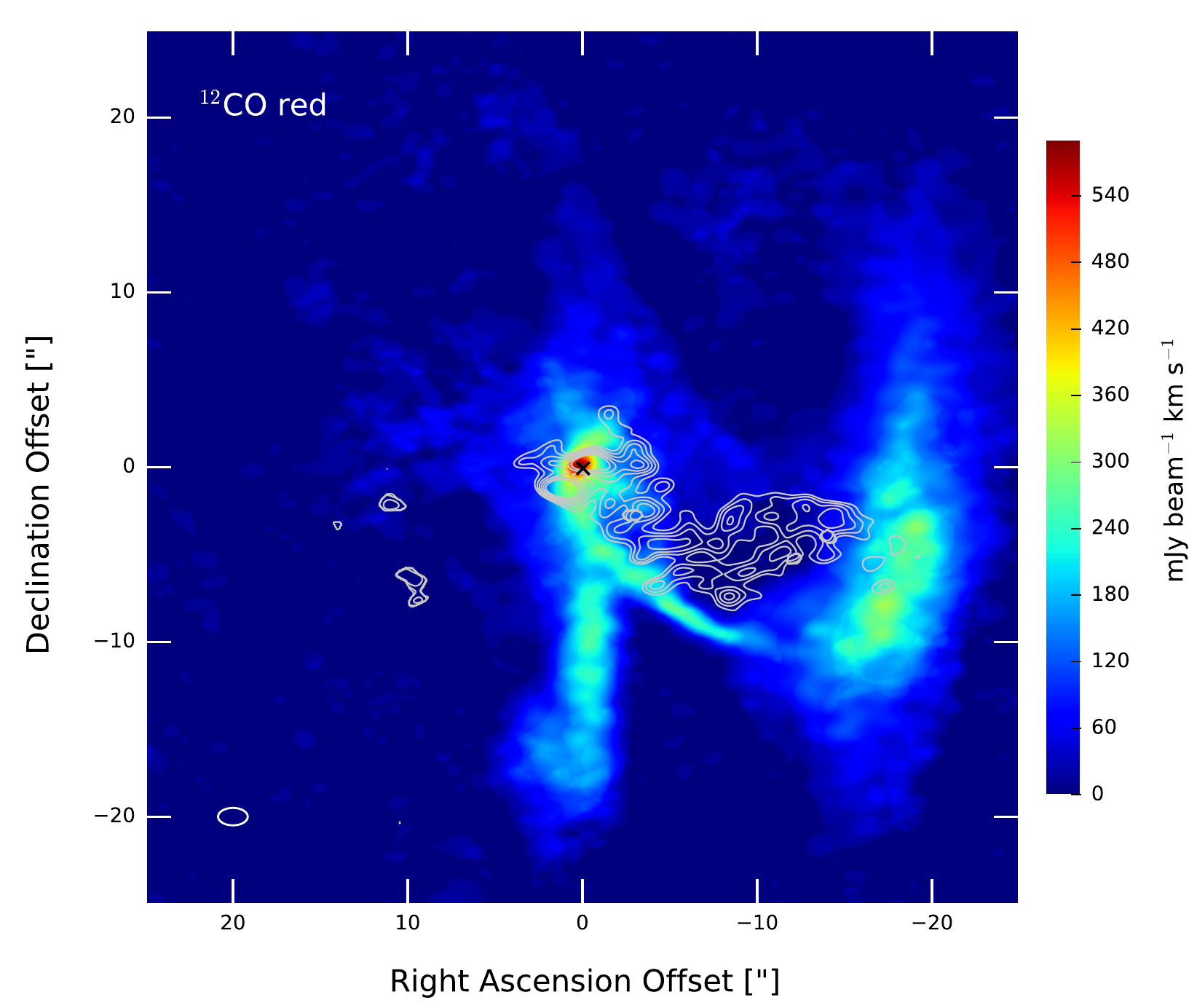}
\caption{ Redshifted C$^{18}$O contours (0.6, 0.7, 0.85, 1.0, 1.1, and 1.2 $\times$ 3$\sigma$) overlaid on the redshifted $^{12}$CO integrated intensity image of V1647 Ori.}
\label{contours_extended_v2}
\end{figure}

\subsection{Dynamics of the V1647 Ori Environment}

\begin{figure*}
\centering
\includegraphics[scale=0.8]{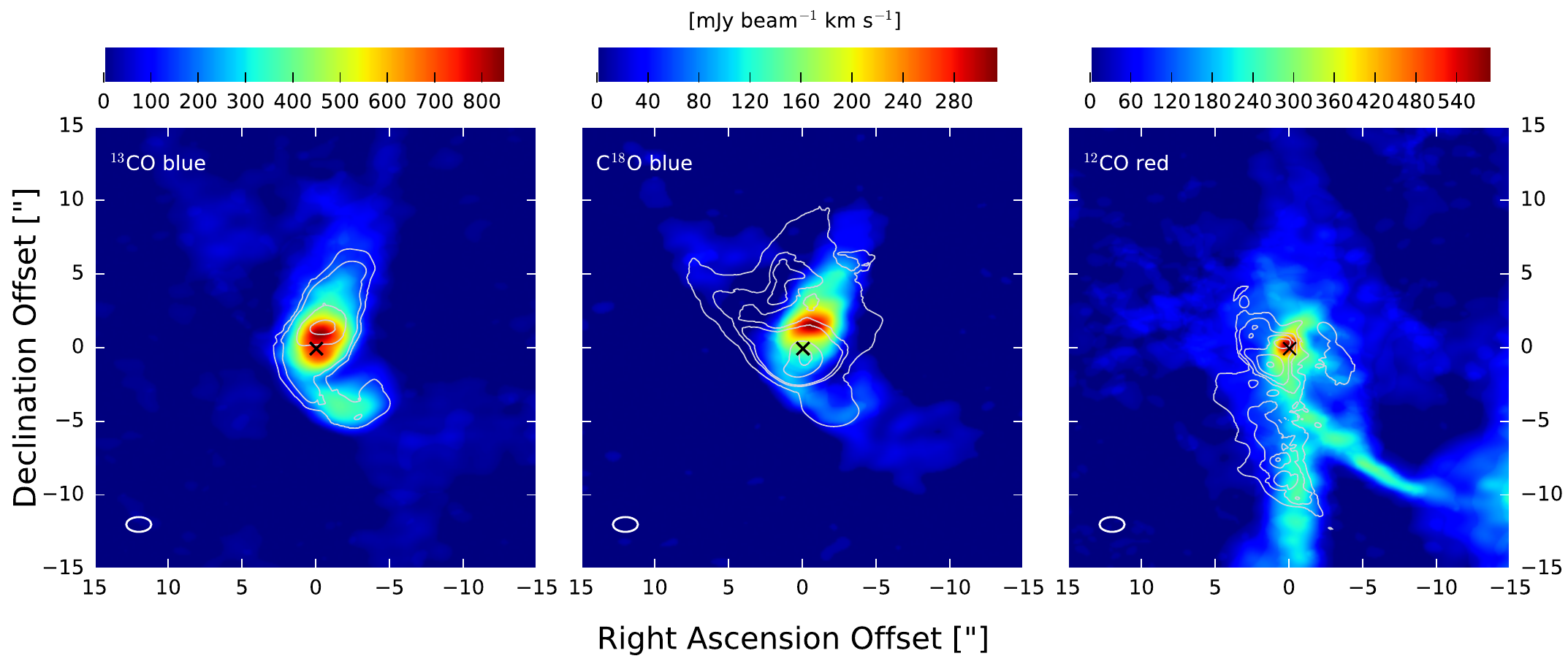}
\caption{ Contours overplotted on blueshifted $^{13}$CO, C$^{18}$O and redshifted $^{12}$CO integrated intensity images of V1647 Ori to highlight outflow/cavity morphology. {\it Left:} Blueshifted C$^{18}$O contours with levels of 1.0, 1.75, 4.0, 6.0, and 8.0 $\times$ 3$\sigma$.  {\it Middle:} Blueshifted $^{12}$CO contours with levels 4.0, 10.0, 12.0, 25.0 and 35.0 $\times$ 3$\sigma$.  {\it Right:} Redshifted $^{13}$CO contours with levels 1.0, 1.5, 2.0, 2.5, and 3.2 $\times$ 3$\sigma$.}
\label{contours_jet}
\end{figure*}

Figure \ref{contours_jet} highlights the various outflows/features identified in the previous sections (labels D and E from Figures \ref{$^{12}$CO_mom0_shifts}, \ref{$^{13}$CO_mom0_shifts}, and \ref{C$^{18}$O_mom0_shifts}).  These features can be summarized in the following way: a $\sim$50$^{\circ}$ collimated outflow at position angle of $\sim$330$^{\circ}$ is detected in blueshifted $^{13}$CO and C$^{18}$O (Figure \ref{contours_jet} {\it left}), a $\sim$100$^{\circ}$ excavated cavity pointing directly north of V1647 Ori is detected in both blueshifted $^{12}$CO and systemic C$^{18}$O and misaligned with the aforementioned collimated outflow  (Figure \ref{contours_jet} {\it center}), and a $\sim$30$^{\circ}$ collimated outflow detected in redshifted $^{12}$CO and $^{13}$CO pointing directly south of V1647 Ori (Figure \ref{contours_jet} {\it right}).  The dynamical age ($\tau_{d}$) of these outflows can be estimated using their physical extent measured from the location of the V1647 Ori disc ($r_\mathrm{outflow}$) and their maximum velocities relative to the velocity of the system (i.e., $\tau_{d}$ =  $\frac{r_\mathrm{outflow}}{v_\mathrm{system}-v_\mathrm{max}}$).  However, these dynamical ages are relatively uncertain given the low inclination of this system ({\it i} = 17$^{\circ}$).  Since each of the two collimated outflows were detected with multiple emission lines, we can estimate two independent dynamical ages for each outflow. These dynamical ages and the values used to determine them are displayed in Table \ref{jet_table} and indicate the outflow identified in blueshifted $^{13}$CO and C$^{18}$O has a dynamical age of $\sim$11,700 years whereas the outflow identified in redshifted $^{12}$CO and $^{13}$CO has a dynamical age of $\sim$17,200 years. Given an uncertainty of $\pm$7 pc for a distance of 414 towards the Orion Nebula \citep{Menten2007}, these dynamical age estimates may be different by $\sim$200 and 300 years, respectively. However, we caution that the dynamical age of the redshifted $^{12}$CO feature is highly uncertain given that it extends to radii larger than the ALMA primary beam. We do not estimate the dynamical age of the $^{12}$CO bowl-like feature pointed directly north because this feature is more likely a cavity rather than an outflow. However, it should be noted that the northern bowl-like emission has a similar blueshifted $^{12}$CO maximum velocity ($v_\mathrm{max}$ = 3) as the collimated redshifted $^{12}$CO feature pointed directly south ($v_\mathrm{max}$ = 2.5).

\setcounter{footnote}{0}
\begin{table*}
\centering
\caption{Dynamical Ages of the Outflows in V1647 Ori}

\begin{tabular}{c c c c c }

\hline
 &  Line & Size & \footnote{1}$v_\mathrm{max}$  & Dynamical Age   \\
 & & [au] & [km s$^{-1}$]  & [years]  \\
\hline
Northern Outflow & $^{13}$CO blueshifted  & {  4500 } & {  1.75}  & {  12,200}  \\
 & C$^{18}$O blueshifted & {  4150 }  & {1.75}  & {11,300}  \\
\hline
Southern Outflow   & $^{12}$CO redshifted  & 9100 & 2.50 & 17,300  \\
 & $^{13}$CO redshifted & 5400 & 1.50 & 17,100  \\
\hline
\end{tabular}

\hspace{4in}\footnotemark[1]{ Maximum velocity relative to a system velocity of 10 km s$^{-1}$.}
\label{jet_table}
\end{table*}

Physical characteristics (e.g., mass, momentum, and kinetic energy) of the molecular gas surrounding the V1647 Ori protostar can be estimated using primary beam corrected $^{12}$CO and $^{13}$CO line emission.  We follow the procedure outlined in \citet[][and ref. therein]{Perez2015} and adopt a $^{12}$CO to H$_{2}$ abundance ratio of 10$^{-4}$ and a $^{12}$CO to $^{13}$CO relative abundance of 62 \citep{Langer1993}.  This analysis assumes a constant excitation temperature of T$_\mathrm{ex}$ = 50 K along the line of sight and a beam filling factor of 1.  We then calculate the outflow mass M =$\sum_{v}$M($x,y,v$),  momentum P = $\sum_{v}$M($x,y,v$) $\times$ $v$, and kinetic energy KE = $\sum_{v}$M($x,y,v$) $\times$ $\frac{1}{2}$$v^{2}$ for each velocity channel within a radius of 10$''$ of V1646 Ori (e.g., see Figure \ref{mom0_all_lines}). This calculation was performed on primary beam-corrected channel maps. Parameters for the blue and redshifted regions of $^{12}$CO and $^{13}$CO are displayed in Table \ref{momentum_table} and correspond to the velocity ranges indicated at the top of Figure \ref{spec_lines} and displayed in Figures \ref{$^{12}$CO_mom0_shifts} and \ref{$^{13}$CO_mom0_shifts}.   Optical depth corrections, typically performed using the ratio of line fluxes between the optically thick $^{12}$CO and optically thin $^{13}$CO \citep{Dunham2014}, are unable to be applied to this dataset because there are few channels where both $^{13}$CO and $^{12}$CO are detected.  Therefore, the $^{12}$CO parameters estimated in this work (Table \ref{momentum_table}) are lower limits, and not necessarily representative of the entire mass of the outflow in $^{12}$CO.

\begin{table*}
\centering
\caption{Physical Properties of the V1647 Ori Outflows}

\begin{tabular}{c c c c c c}

\hline
Molecule &  Side & Column Density & Mass  & Momentum  & Energy  \\
 & & [10$^{20}$ m$^{-2}$] & [$10^{-3}$ $M_{\odot}$]  & [$10^{-3}$ $M_{\odot}$ km s$^{-1}$] & [$10^{40}$ erg] \\
\hline
$^{12}$CO & Blueshifted  & {  9.8} & {  22.6}  & {  34.7} & {60.3} \\
 & Redshifted & {  5.5}  & {  11.1}  & {  20.5} & {  39.8} \\
 & Total & {  15.3} & {  33.7}  & {  55.2} & {  100.1} \\
\hline
$^{13}$CO   & Blueshifted  &  5.4 & 11.7 & 9.0 & 7.4 \\
 & Redshifted  & 2.1 & 3.6 & 5.2 & 7.9 \\
 & Total & 7.5 &  15.3 & 14.2 & 15.3  \\
\hline
\end{tabular}

\label{momentum_table}
\end{table*}

\subsection{Optical Reflection Nebula (McNeil's Nebula)}

\begin{figure*}
\begin{minipage}{0.46\textwidth}
\includegraphics[scale=0.77]{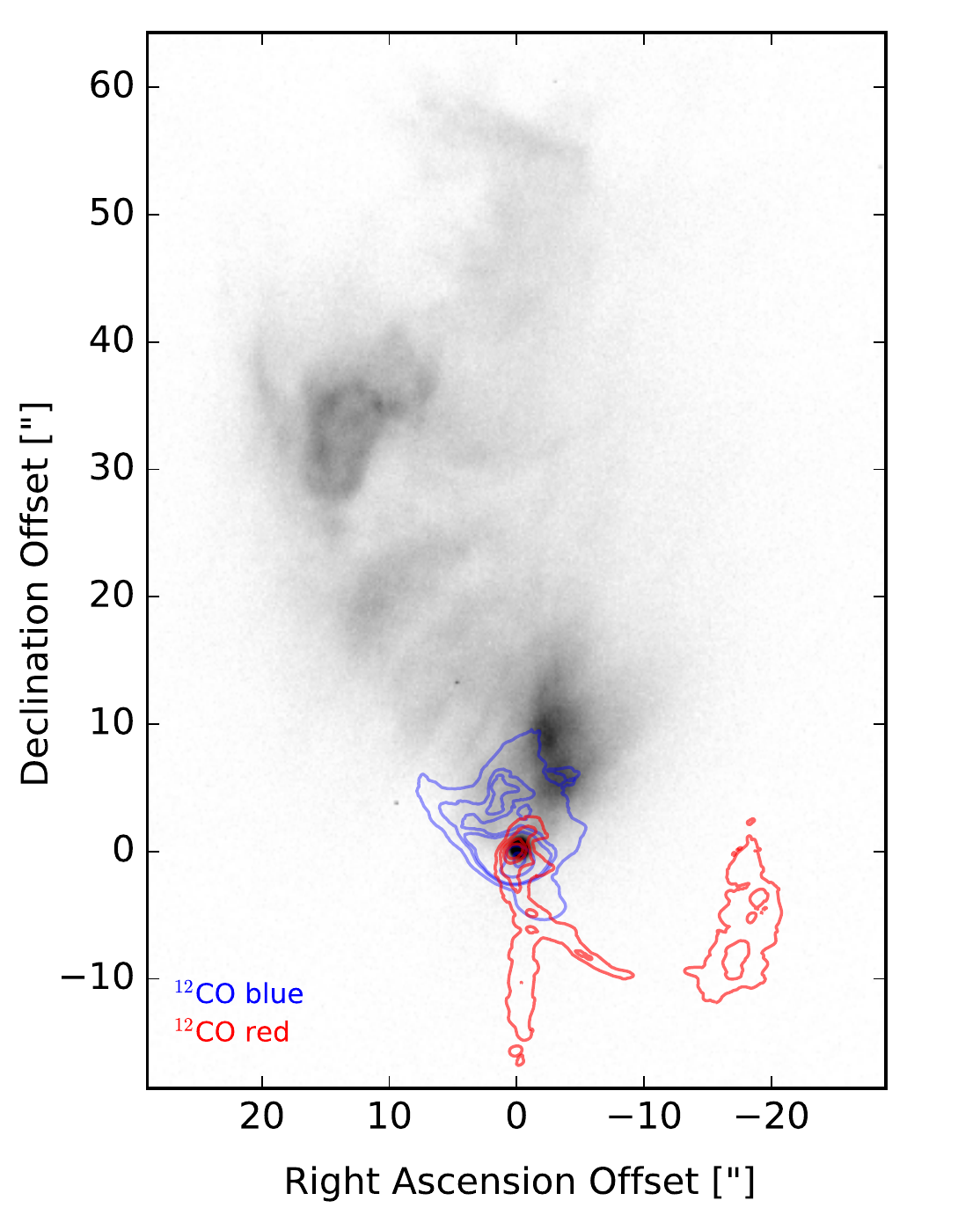}
\end{minipage}
\begin{minipage}{0.25\textwidth}
\vspace{-6mm}
\includegraphics[scale=0.35]{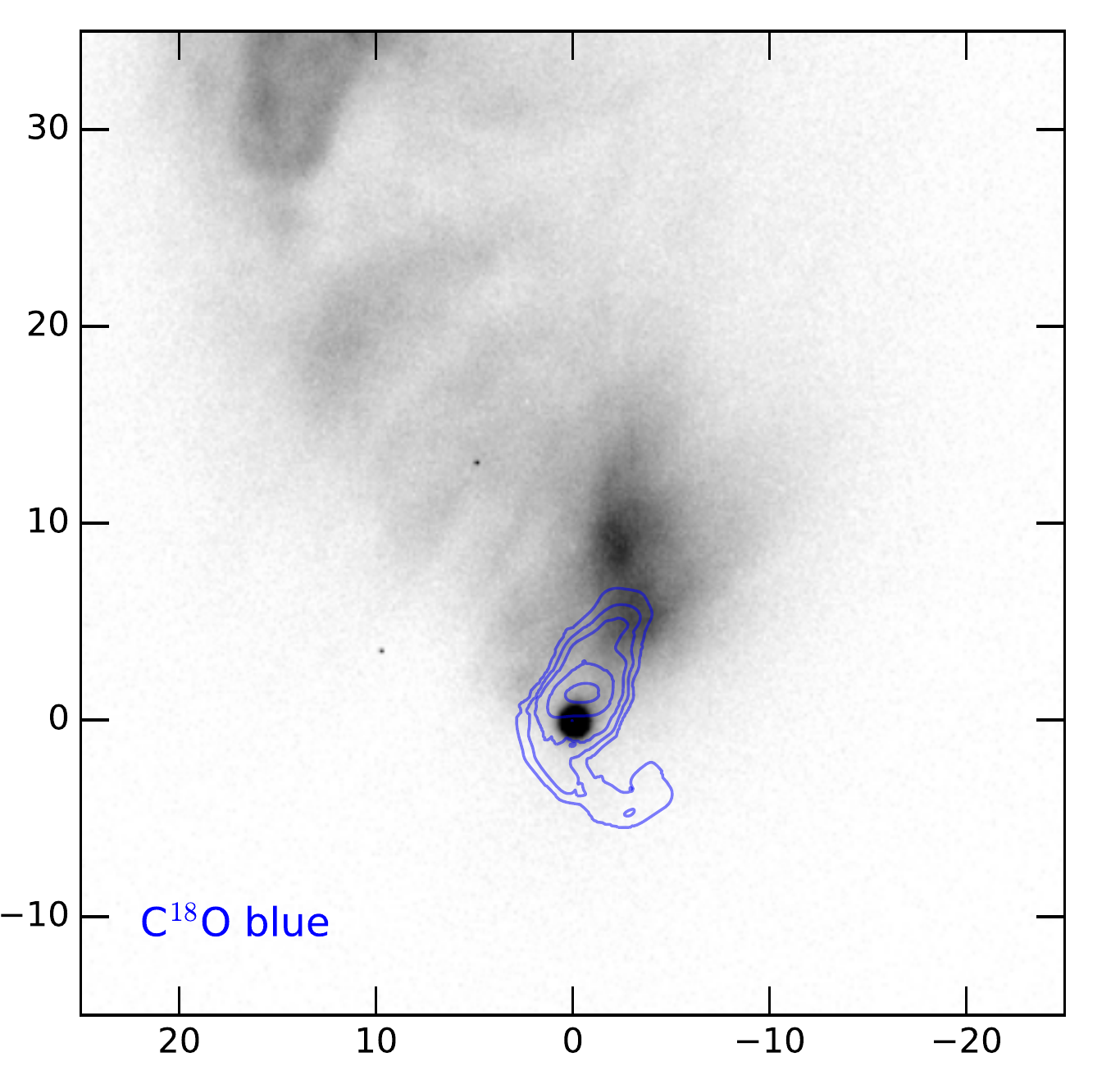}
\includegraphics[scale=0.35]{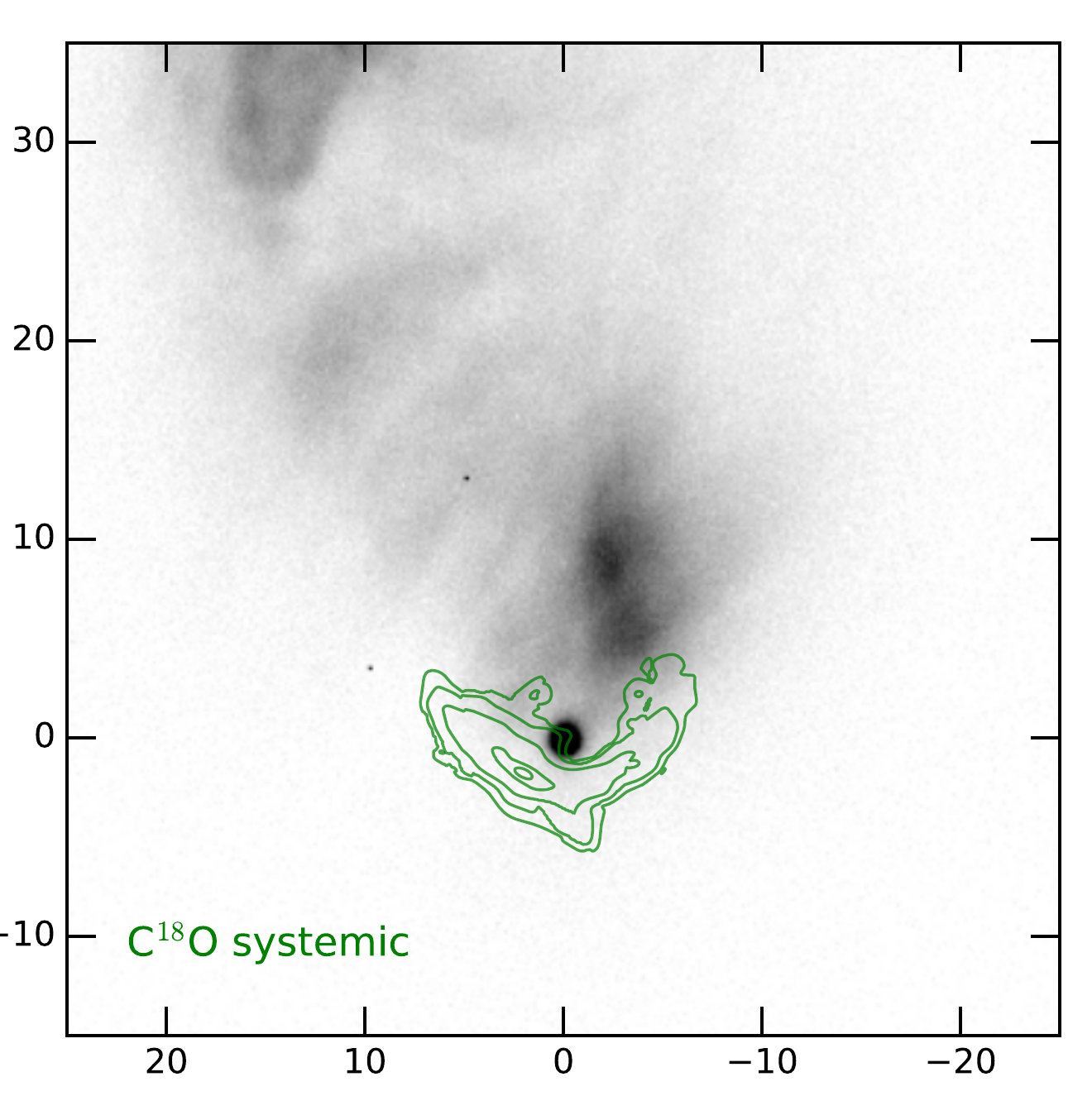}
\end{minipage}
\caption{ Optical Gemini GMOS r band observation of V1647 Ori and its reflection nebula taken during the 2008 outburst.  Overlaid are integrated intensity image contours of $^{12}$CO, $^{13}$CO, and C$^{18}$O as indicated in the lower left of each panel. North is up and east is left  }
\label{optical_contours}
\end{figure*}

The $\sim$1$'$ diameter optical r-band reflection nebula to the north of V1647 Ori is displayed in Figure \ref{optical_contours}.  These data were taken during the 2008 outburst and display a generally diffuse nebulosity of dust with a clump of material to the north west near RA and Dec offset [-4, 10].  Overlays of the $^{12}$CO, $^{13}$CO, and C$^{18}$O contours on the optical data indicate the millimeter gas emission is well aligned with the base of the reflection nebula.  Moreover, the collimated outflow observed in $^{13}$CO and C$^{18}$O appear to be traveling in the direction of the bright dust clump observable at optical wavelengths.

\section{Discussion}

\subsection{The Circumstellar Disc of V1647 and Potential Mechanism for Accretion Outbursts}

The resolved continuum image of the V1647 Ori circumstellar disc (Figure \ref{continuum}) does not display clear evidence of spirals or clumps associated with disc fragmentation or perturbations from a potential binary companion.  Therefore, it is unlikely there are any binary or massive substellar companions $\gtrsim$ 40 au from the central star capable of producing the intense outbursts observed in V1647 Ori. However, the potential non-gaussian radial brightness profile identifiable at large radii in Figure \ref{continuum} suggests some amount of asymmetry and warrants a more detailed investigation than that performed for the continuum data and presented here.   While the circumstellar disc is clearly detected in the dust continuum, it is less evident in the molecular line data.  Molecular emission spatially coincident with the location of the disc is identifiable in the redshifted $^{12}$CO integrated intensity image (Figure \ref{$^{12}$CO_mom0_shifts}) and is potentially detected but unresolved from extended cloud/outflow emission at blueshifted $^{12}$CO and $^{13}$CO and redshifted $^{13}$CO and C$^{18}$O. Such emission in these cases may not be from the disc but instead could be from a small region surrounding the disc.  Moreover, gaseous emission from the disc was not detected in any of the emission lines at the systemic velocities (e.g., see Figure \ref{spec_lines} top and Appendix).  Given the inconsistency of its detection, we were unable to search for signatures of Keplerian rotation.  The non-detection of the disc at systemic velocities is likely due to the deeply embedded nature of this system which is consistent with its high extinction at optical \citep[A$_{V}$ $= 8-19$;][and references therein]{Audard2014} and X-ray \citep[N$_{H}$ $\sim$ 4 $\times$ 10$^{22}$ cm$^{-2}$;][]{Teets2011} wavelengths.

The radius of the disc derived from the continuum emission (Table \ref{continuum_table}) is similar to that of the namesake of the FUor class \citep[FU Ori;][]{Hales2015} as well as HBC 494 and V2775 Ori \citep{Ruiz-Rodriguez2017a,Zurlo2017}.  However, compared with V883 Ori, the V1647 Ori disc is a factor of $\sim$2 smaller and its star/disc bolometric luminosity is a factor of 10 fainter \citep{Audard2014}.  The V1647 Ori disc mass derived from the continuum is the least massive compared with the other FUor objects presented in this series. This suggests objects that have been classified strictly as FUors may be more massive than EXor objects where V1647 Ori is in the middle of this classification as it displays outbursts indicative of both classifications.

\subsection{Excavated Cavities and Outflows}

The molecular emission line data indicate that the star-forming environment surrounding V1647 Ori is both structured and complex.  Several features indicating the three-dimensional structure of this system are readily identifiable and show that the protostar and disc system are still deeply embedded in its parent molecular cloud. The orientation of the system can be further constrained when considering the direction of the blueshifted outflow (label E) coincident with northern reflection nebula (Figure \ref{optical_contours}).  The dust cavity appears in the north at optical wavelengths because it is preferentially forward scattering emission from its illuminating source (i.e., V1647 Ori during outburst).  This indicates that the cavity is located between V1647 Ori and our line of sight which is consistent with the cavity detection in blueshifted $^{12}$CO emission (label A).  The co-alignment of this excavated northern cavity with the outflow at PA$\sim$330$^{\circ}$ is further substantiated by the location of the bright clumps visible during outburst in the optical reflection nebula since they are aligned with direction of the collimated outflow (Figure \ref{optical_contours}). This orientation also explains the absence of a reflection nebula at optical wavelengths to the south of V1647 Ori even though a southern cavity is present (Figure \ref{optical_contours}; i.e., the southern cavity is pointed away from us and any reflected light will be backscattered and too faint to detect in shallow observations).

Given the appearance of its complex environment, the strongest constraints on the orientation of the V1647 Ori system is measured by the gaussian fit to the continuum data which indicates the disc is almost face-on ({\it i} = 17$^{\circ}$) with a position angle of 109$^{\circ}$. This disc inclination is in good agreement with previous estimates of {\it i} $<$30$^{\circ}$ \citep{Andrews2004,Rettig2005} but is not consistent with the stellar inclination modeled from periodic X-ray emission generated by rotating accretion hot spots \citep[{\it i} $\sim$ 68$^{\circ}$;][]{Hamaguchi2012}. The measured position angle of the disc (109$^{\circ}$ $\pm$19) appears to align more with the outflow at PA$\sim$330$^{\circ}$ (label E) than the southern collimated outflow at PA$\sim$180$^{\circ}$ (label D) assuming that outflows travel perpendicular to the plane of the disc and the disc is inclined such that the northern extended emission is blueshifted towards our line of sight.  Since the disc orientation is a strong constraint on the system as a whole, our interpretation of the large scale (r $\gtrsim$ 1-3$''$) emission features (e.g., collimated outflows) assumes that these features were formed from material traveling roughly perpendicular to the circumstellar disc of V1647 Ori at the time of their respective outflow triggering events.  Such an assumption is reasonable given our current understanding of launching mechanism of material during protostellar evolution \citep[][and references therein]{Frank2014}.

The blueshifted $^{12}$CO emission whose bowl-like shape conforms well to the base of the optical reflection nebula (Figure \ref{optical_contours}) is seemingly misaligned with the aforementioned blueshifted $^{13}$CO and C$^{18}$O collimated feature (Figure \ref{contours_jet}, center).  Moreover, the non-detection in $^{13}$CO of this $^{12}$CO blueshifted bowl-like shape suggests that there are no active wide-angle outflows at this location and that the $^{12}$CO is probing the temperature of the walls of the excavated cavity.  The misalignment between the collimated southern outflow (Figures \ref{$^{12}$CO_mom0_shifts}, \ref{$^{13}$CO_mom0_shifts}, label D) and the collimated northern outflow at PA$\sim$330$^{\circ}$ (Figures \ref{$^{13}$CO_mom0_shifts}, \ref{C$^{18}$O_mom0_shifts}, label E) may be the result of a re-orientation or precession of the mass-launching area (i.e., the orientation of the circumstellar disc may have changed from one outburst to another resulting in two outflows orientated with different position angles).  The difference in position angle and the dynamical age between these collimated outflows (Table \ref{jet_table}) suggest each outflow was likely the result of a specific mass-accretion event with the southern outflow occurring $\sim$5000 years before the north-west outflow.  Given that a radical change in disc position angle is unlikely, these collimated outflows were probably launched from opposite sides of the circumstellar disc during their respective mass accretion events.

While some asymmetric jets, like those detected from Herbig-Haro objects, appear to be shaped by their local star-forming environment \cite{Bally2001}, others systems exhibit clear evidence that the jet-launching region (i.e., the region perpendicular to the plane of the disc) has precessed over time resulting in outflows with asymmetric and misaligned features \citep{Shepherd2000,Su2007,Cunningham2009b,Beltran2016}.  While the amount of precession over the lifetime of the outflow varies from case to case, it appears that jet precession of more than a few degrees may be rare \citep{Frank2014}.  In cases where the degree of disc precession is high (like in the case of V1647 Ori), several mechanisms have been proposed capable of significantly re-orienting the outflow/jet: (a) radiative-induced disc warping where a jet, presumably oriented perpendicular to a disc, precesses in response to a disc warp which is generated in cases of high-incident radiation \citep{Armitage1997}; (b) tidal interaction from a companion in a non-coplanar orbit truncating and/or distorting the disc \citep{Lubow2000, Terquem1999} ; or (c) anisotropic accretion events where an impact/merging of material onto the disc during anisotropic accretion can change the orientation of the disc angular momentum vector resulting in a net torque in the rotation of the disc \citep{Shepherd2000,Kraus2006}.

Estimating the critical radius at which the disc becomes unstable to warping following equation 5 in \citet{Armitage1997}, the assumption that $\eta$=1 in \citet{Shepherd2000}, and the physical characteristics of V1647 Ori during outburst (M = 0.8 $M_{\odot}$, $\overset{.}{M}$$_\mathrm{acc}$= 4 $\times$ $10^{-6}$ $M_{\odot}$ yr$^{-1}$, $L = 44 L_{\odot}$),  $R_\mathrm{crit}$ = 2 $\times$ 10$^{6}$ au.  Given that this radius is orders of magnitude larger than the radius of the V1647 Ori disc, it is unlikely a disc warp is the origin of its misaligned outflows. However, it should be noted that the stellar parameters during the outburst event that produced these outflows ($\sim$14,000 years ago; Table \ref{jet_table}) are not necessarily the same as those measured during the 2003 outburst.  As noted in Section 4.1, the non-detection of any companions to V1647 Ori at radii > 40 au likely rules out tidal interactions via a companion causing disc precession unless such a companion exists at smaller radii and has not disrupted the V1647 Ori disc in such a way as to be evident from the continuum observations.  Therefore, given the intrinsic connection between accretion and FUor/EXor outbursts, it appears that significant mass-loading (e.g., from a large clump of envelope material or a companion) during anisotropic accretion may be the most likely scenario to have caused the disc to precess between each outflow-generating outburst in V1647 Ori.  

If this anisotropic accretion scenario is true in the case of V1647 Ori, one might expect more FUor/EXor type objects to display multiple misaligned outflows. However, there appears to be little evidence for this, at least in the case of the other FUors studied as part of this series \citep{Zurlo2017,Ruiz-Rodriguez2017a,Ruiz-Rodriguez2017b}. Given the complex geometry of the V1647 Ori star-forming environment, it may be that other FUor/EXor objects underwent smoother mass-accretion outbursts which did not result in disc precession or simply had fewer large-scale mass accretion events.  Z CMa, a binary FUor object that also exhibits properties of an EXor, displays two slightly misaligned outflows where a wiggling shape of one of the outflows may be generated via precession from an unseen additional companion \citep{Whelan2010,Antoniucci2016}. Given that Z CMa and V1647 Ori represent a rare class of objects that exhibit properties of both FUor and EXor objects, we speculate that disc precession may be related to this intermediate classification.

While the disc precession argument seems the most plausible, other scenarios could be invoked to describe the complicated geometry of these two collimated emission features. The two outflows could originate from separate binary members ( e.g., each with their own outflow) if V1647 Ori has an undetected companion.   In this case, outflow misalignment may be a signature that the potential binary formation of V1647 Ori resulted from turbulent fragmentation \citep{Offner2016}.  However, this scenario generally considers binaries with separations much larger than what would likely be the case for V1647 Ori ($\lesssim$ 40 au). Another scenario alternative to changes in disc orientation is the case that the outflows instead are located at different position angles as a result of a dense material within the molecular cloud shaping the outflow.  V1647 Ori clearly resides in a complex and embedded environment and the outflow structure could simply be following a path of least resistance due to some unobserved molecular cloud component whose mass would be difficult to determine without knowledge of the pre-outflow geometry.  This scenario would also affect the dynamical age calculations if the outflowing material does not travel in a straight line.  Another explanation alternative to disc precession could be that the collimated features observed are actually inflows and not outflows. This interpretation would require our understanding of the system geometry to be  reversed (i.e., the blueshifted outflow situated between V1647 Ori and our line of sight would instead be behind V1647 Ori and traveling towards the system presumably in a large accretion stream).  However, this scenario is unlikely to be true given the spatial correlation between the blueshifted $^{13}$CO and C$^{18}$O emission with the northern cavity of the optical reflection nebula which independently indicates this northern region is between V1647 Ori and our line of sight and thus, the emission feature is more likely an outflow.

\subsection{The Shaping Potential of Dense Molecular Gas Surrounding V1647 Ori}

Given the structure of the red and blueshifted molecular line emitting material compared to that of the systemic C$^{18}$O emission (Figure \ref{C$^{18}$O_system_contours}), it is evident the dense material in which the protostar is embedded influences the overall structure of the system. While it is not clear whether this dense material can result in two misaligned outflows (Section 4.2), it is still evident that the blueshifted $^{12}$CO bowl-like shape is surrounded by the dense C$^{18}$O material of similar shape (Figure \ref{C$^{18}$O_system_contours} left).  This is in good agreement with the location of the wall of the excavated cavity as seen in $^{12}$CO, suggesting the shaping mechanism that produced this cavity compressed the low-density ambient molecular material until it collided with the dense material as probed by C$^{18}$O. Moreover, the western arm of the systemic C$^{18}$O material appears to have cut off or 'pinched' the material associated with blueshifted $^{12}$CO and $^{13}$CO material $\sim$3$''$ southwest of the V1647 Ori disc (label B). The outflowing blueshifted material appears to travel around the dense material at systemic velocities as it may be more energetically favorable. If this interpretation is true then the location of the dense systemic C$^{18}$O material must be between our line of sight and the V1647 Ori disc (i.e., in the foreground but still part of the V1647 Ori star-forming environment). This interpretation suggests that the local dense gas plays a large role in the shaping of cavities and molecular outflows during this stage of pre-MS stellar evolution.

A similar shaping scenario may be taking place where the redshifted C$^{18}$O emission appears along the border of the 'arm' seemingly connecting the V1647 Ori system to the large unresolved feature $\sim$20$''$ (8300 au) to the west (Figure \ref{contours_extended_v2}).  This large $\sim$40$''$ (17,000 au) column of material is also detected in a few channels of $^{13}$CO emission traveling at the same velocity (Figure 2 in the appendix).  The source of this structure is unclear and it may just be leftover cloud material that either has yet to form a star or does not meet the physical requirements to do so.  However, we caution that this feature may be an artifact from emission outside the primary beam.   The closest pre-MS star (in projection) to this structure (2MASS J05461162-0006279) is located $\sim$ 7$''$ (3150 au) from this extended structure's south-western edge.  It is identified as a Class II object in the L1630 star-forming region based on its infrared-excess and detection in X-rays and has a moderate absorption of N$_{H}$ = 5 $\times$ 10$^{21}$ cm$^{-2}$ \citep{Principe2014}.

\subsection{Outflow Dynamics and a Comparison with Other FUors}

The complex structures identified in the V1647 Ori at spatial scales of $\sim$1.0$''$  can be compared with the other similarly structured FUor/EXor type objects presented in this series \citep[e.g.,][]{Zurlo2017,Ruiz-Rodriguez2017a,Ruiz-Rodriguez2017b}.  Of the eight FUor/EXor objects observed, only V2775 Ori, V883 Ori, HBC 494 and V1647 Ori display significant amounts of extended $^{12}$CO, $^{13}$CO and/or C$^{18}$O emission.  Given the high A$_V$, N$_H$, complex C$^{18}$O environment, and ambiguous detection of its gaseous circumstellar disc, V1647 Ori appears to be in the earliest stage of stellar evolution compared with the other FUor objects presented.  In particular, the structures displayed in these ALMA observations of V1647 Ori could represent the progenitor stages of those of V883 Ori and V2775 Ori given their similar inclination (and thus similar viewing angle). Similar to the case of V2775 Ori \citep[Figure 7 in][]{Zurlo2017}, V1647 Ori also exhibits a small outflow opening angle ($\sim$30$^{\circ}$) in $^{13}$CO, C$^{18}$O, further indicating its primordial nature if opening angle is a correlation to age. In contrast, HBC 494 and V883 Ori \citep[Figures 4 and 2, respectively, in][]{Ruiz-Rodriguez2017b,Ruiz-Rodriguez2017a} have much larger opening angles of $\sim$150$^{\circ}$.  Almost counterintuitively, the outflows of V1647 Ori have the oldest dynamical ages (11,700 and 17,200 years) when compared to that of V2775 Ori \citep[1600-2600 years][]{Zurlo2017}, HBC 494 \citep[5400 years][]{Ruiz-Rodriguez2017b} and V883 Ori \citep[10,000 years][]{Ruiz-Rodriguez2017a}.  These dynamical ages are very small given the timescale of pre-MS stellar evolution with respect to disc dissipation (e.g., $\sim$1-3 Myr).  Therefore the older dynamical timescale of V1647 Ori's outflows do not necessarily mean it is inconsistent with the interpretation here that this system is less evolved than its other FUor counterparts.  It could simply mean that whatever mechanism initiates these outbursts may have started earlier in V1647 Ori.  Given the complex environment surrounding V1647 Ori, it may indicate that inhomogeneous patches of envelope material could have prompted more frequent or earlier accretion outbursts.

 The outflow mass surrounding V1647 Ori (as indicated in Section 3.6) appears similar to those obtained for other Class 0-II pre-MS stars with outflows \citep{Arce2006} as well as the small sample of FUor type objects whose outflows have been measured \citep{Evans1994}.  However, when comparing the outflow velocities associated with these objects, it is clear that the outflows surrounding V1647 Ori are quite slow ($\sim$1-2 km s$^{-1}$).  Given the similar sensitivity and resolution of the other FUors detected as part of this series, we compare the results of V1647 Ori directly with those of V2775 Ori, HBC 494, and V883 Ori.  Of these sources, V1647 Ori has the least massive circumstellar disc and is the only source considered both an FUor and EXor type object based on its unusual outburst frequency.  V1647 Ori is the most similar in disc mass and inclination to V2775 Ori \citep[M$_\mathrm{disc}$ $\sim$ 530 $M_{E}$, {\it i} $\sim$ 14$ ^{\circ}$;][]{Zurlo2017,Cieza2017} however their local environments are quite distinct.  While V2775 Ori displays an impressive ring-like structure in its blue and red shifted emission \citep[ Figure 3 in][]{Zurlo2017}, there is little evidence of complex structure surrounding the large-scale environment of the protostar indicating that this system may be more evolved than V1647 Ori.  Moreover, V2775 Ori exhibits a redshifted $^{12}$CO outflow velocity of 4.4 km s$^{-1}$, a factor of $\sim$2 faster than those of V1647 Ori, potentially indicative of a emptier environment surrounding the star where the outflow doesn't slow down from interactions with surrounding material. A comparison of outflow mass and energy \citep[e.g. Table 3 in][]{Zurlo2017} indicates the outflow of V1647 Ori as probed by $^{12}$CO and $^{13}$CO is more massive than those in V2775 Ori by a factor of 3 and 15, respectively.   A comparison of V1647 Ori with that of HBC 494 and V883 Ori, both of which have more massive and larger circumstellar discs, indicates the outflow mass of V1647 Ori is significantly smaller \citep{Ruiz-Rodriguez2017b,Ruiz-Rodriguez2017a}.  Both these systems exhibit slow outflow speeds similar to or smaller than V1647 Ori possibly indicating they too still are interacting with or have been influenced by surrounding material.

\section{Summary and Conclusions}

We have presented ALMA  observations of the V1647 Ori continuum at a resolution of 0.2$''$ as well as molecular line emission of $^{12}$CO,$^{13}$CO and C$^{18}$O at a resolution of $\sim$1$''$. No emission line features were detected in the 0.2$''$ dataset indicating that structure surrounding V1647 Ori are fairly uniform on large scales. These observations are presented as part of a series and we compare these results of V1647 Ori to that of other FUor type objects with extended emission (e.g., V2775 Ori, HBC 494, and V883 Ori).  We have identified a complex yet structured system surrounding the V1647 Ori protostar and circumstellar disc and are able to draw several conclusions:

\begin{description}

\item[$\bullet$] The circumstellar disc of V1647 Ori is resolved and has a nearly face-on inclination ({\it i} = 17$^{\circ}$), a radius of 40 au and a total disc mass of $\sim$0.1 M$_{\odot}$.   The continuum resolved disc does not display any disc structures (e.g., spirals and/or fragmentation) associated with nearby interacting companions (r $\gtrsim$ 40 au) and therefore such a mechanism, under these constraints, is unlikely to be responsible for the various observed accretion outbursts.  The mass of the disc is less than the other FUor type objects presented in this series may be related to its classification as both an FUor and EXor type object.

\item[$\bullet$] Blueshifted $^{12}$CO molecular line emission spatially coincident with the V1647 Ori optical reflection nebula (McNeil's Nebula) and a non-detection of this feature in $^{13}$CO and C$^{18}$O emission confirm that this structure is a cavity excavated during a previous accretion outburst.  A collimated $^{13}$CO and C$^{18}$O emission feature aligns with clumps in the nebula identified during the recent multiwavelength outburst.

\item[$\bullet$] Dense molecular material as probed via C$^{18}$O emission traveling at a systemic velocity of $\sim$10 km s$^{-1}$ is spatially coincident with the base of the northern excavated cavity and likely shaped the outflow material identified in blueshifted $^{12}$CO, $^{13}$CO, and C$^{18}$O emission.

\item[$\bullet$] Two distinct collimated ($\sim$30-50$^{\circ}$) outflows with position angles of 330$^{\circ}$ and 180$^{\circ}$ have been detected with dynamical ages of $\sim$11,700 and $\sim$17,100 years, respectively.  These outflows are likely the result of two distinct mass accretion events where outflows were launched from opposite sides of a disc that has changed orientation over the last $\sim$5000 years from previous anisotropic accretion events.

\item[$\bullet$] Compared to the other FUor objects included in our series, V1647 Ori displays the most complex large-scale structures likely the result of its comparatively early stage of stellar evolution. A comparison of outflow parameters supports this notion and also indicates that the outflow velocities of these FUor and FUor/EXor type objects ($\sim$0.5 - 4 km s$^{-1}$) are relatively small compared to those observed in typical pre-MS stars and predicted by simulations ($v$$\sim$10-30 km s$^{-1}$). The complex star-forming environment surrounding V1647 Ori may result in more frequent yet still extreme outburst events which justify its classification as both an FUor and EXor type object.

\item[$\bullet$] A large ($\sim$17,000 au) unresolved structure is identified $\sim$8300 au from V1647 Ori.  This structure is not associated with any obvious source but is seemly connected to V1647 Ori by an 'arm' of molecular material as probed with redshifted $^{12}$CO emission.

\end{description}

\section*{Acknowledgements}

This research was supported by CONICYT-FONDECYT awards (3150550, 1171246) and support from the Millennium Science Initiative (Chilean Ministry of Economy; grant Nucleus RC 130007). This paper makes use of the following ALMA data: ADS/JAO.ALMA \#2013.1.00710.S . ALMA is a partnership of ESO (representing its member states), NSF (USA) and NINS (Japan), together with NRC (Canada), NSC and ASIAA (Taiwan), and KASI (Republic of Korea), in cooperation with the Republic of Chile. The Joint ALMA Observatory is operated by ESO, AUI/NRAO and NAOJ. The National Radio Astronomy Observatory is a facility of the National Science Foundation operated under cooperative agreement by Associated Universities, Inc.. D. P. acknowledges funding by the National Aeronautics and Space Administration through Chandra Award Number GO6-17013A. K.M acknowledges funding by the Joint Committee of ESO/Government of Chile, and funding by the Science and Technology Foundation of Portugal (FCT), grant No. IF/00194/2015. H.C. acknowledges support from the Spanish Ministerio de Econom\'ia y Competitividad under grant AYA 2014-55840-P. J.J.T acknowledges support from the University of Oklahoma, the Homer L. Dodge endowed chair, and grant 639.041.439 from the Netherlands Organisation for Scientific Research (NWO). The authors would like to thank the referee whose comments and suggestions increased the quality of this work.






\bibliography{principe_YSO_v2017_hyperref}

\begin{thebibliography}{60}
\expandafter\ifx\csname natexlab\endcsname\relax\def\natexlab#1{#1}\fi
\expandafter\ifx\csname href\endcsname\relax
  \def\href#1#2{}\fi
\expandafter\ifx\csname urllinklabel\endcsname\relax
  \def\urllinklabel{[LINK]}\fi
\expandafter\ifx\csname adsurllinklabel\endcsname\relax
  \def\adsurllinklabel{[ADS]}\fi

\bibitem[{{{\'A}brah{\'a}m} {et~al.}(2006){{\'A}brah{\'a}m}, {Mosoni},
  {Henning}, {K{\'o}sp{\'a}l}, {Leinert}, {Quanz}, \& {Ratzka}}]{Abraham2006}
{{\'A}brah{\'a}m}, P., {Mosoni}, L., {Henning}, T., {K{\'o}sp{\'a}l}, {\'A}.,
  {Leinert}, C., {Quanz}, S.~P., \& {Ratzka}, T. 2006, \aap, 449, L13
 \href{http://adsabs.harvard.edu/abs/2006A%26A...449L..13A}{\adsurllinklabel}

\bibitem[{{Acosta-Pulido} {et~al.}(2007){Acosta-Pulido}, {Kun},
  {{\'A}brah{\'a}m}, {K{\'o}sp{\'a}l}, {Csizmadia}, {Kiss}, {Mo{\'o}r},
  {Szabados}, {Benk{\H o}}, {Barrena Delgado}, {Charcos-Llorens}, {Eredics},
  {Kiss}, {Manchado}, {R{\'a}cz}, {Ramos Almeida}, {Sz{\'e}kely}, \&
  {Vidal-N{\'u}{\~n}ez}}]{Acosta-Pulido2007}
{Acosta-Pulido}, J.~A., {Kun}, M., {{\'A}brah{\'a}m}, P., {K{\'o}sp{\'a}l},
  {\'A}., {Csizmadia}, S., {Kiss}, L.~L., {Mo{\'o}r}, A., {Szabados}, L.,
  {Benk{\H o}}, J.~M., {Barrena Delgado}, R., {Charcos-Llorens}, M., {Eredics},
  M., {Kiss}, Z.~T., {Manchado}, A., {R{\'a}cz}, M., {Ramos Almeida}, C.,
  {Sz{\'e}kely}, P., \& {Vidal-N{\'u}{\~n}ez}, M.~J. 2007, \aj, 133, 2020
 \href{http://adsabs.harvard.edu/abs/2007AJ....133.2020A}{\adsurllinklabel}

\bibitem[{{Andrews} {et~al.}(2004){Andrews}, {Rothberg}, \&
  {Simon}}]{Andrews2004}
{Andrews}, S.~M., {Rothberg}, B., \& {Simon}, T. 2004, \apjl, 610, L45
 \href{http://adsabs.harvard.edu/abs/2004ApJ...610L..45A}{\adsurllinklabel}

\bibitem[{{Andrews} \& {Williams}(2005)}]{Andrews2005}
{Andrews}, S.~M. \& {Williams}, J.~P. 2005, \apj, 631, 1134
 \href{http://adsabs.harvard.edu/abs/2005ApJ...631.1134A}{\adsurllinklabel}

\bibitem[{{Antoniucci} {et~al.}(2016){Antoniucci}, {Podio}, {Nisini},
  {Bacciotti}, {Lagadec}, {Sissa}, {La Camera}, {Giannini}, {Schmid},
  {Gratton}, {Turatto}, {Desidera}, {Bonnefoy}, {Chauvin}, {Dougados},
  {Bazzon}, {Thalmann}, \& {Langlois}}]{Antoniucci2016}
{Antoniucci}, S., {Podio}, L., {Nisini}, B., {Bacciotti}, F., {Lagadec}, E.,
  {Sissa}, E., {La Camera}, A., {Giannini}, T., {Schmid}, H.~M., {Gratton}, R.,
  {Turatto}, M., {Desidera}, S., {Bonnefoy}, M., {Chauvin}, G., {Dougados}, C.,
  {Bazzon}, A., {Thalmann}, C., \& {Langlois}, M. 2016, \aap, 593, L13
 \href{http://adsabs.harvard.edu/abs/2016A%26A...593L..13A}{\adsurllinklabel}

\bibitem[{{Arce} \& {Sargent}(2006)}]{Arce2006}
{Arce}, H.~G. \& {Sargent}, A.~I. 2006, \apj, 646, 1070
 \href{http://adsabs.harvard.edu/abs/2006ApJ...646.1070A}{\adsurllinklabel}

\bibitem[{{Armitage} {et~al.}(2001){Armitage}, {Livio}, \&
  {Pringle}}]{Armitage2001}
{Armitage}, P.~J., {Livio}, M., \& {Pringle}, J.~E. 2001, \mnras, 324, 705
 \href{http://adsabs.harvard.edu/abs/2001MNRAS.324..705A}{\adsurllinklabel}

\bibitem[{{Armitage} \& {Pringle}(1997)}]{Armitage1997}
{Armitage}, P.~J. \& {Pringle}, J.~E. 1997, \apjl, 488, L47
 \href{http://adsabs.harvard.edu/abs/1997ApJ...488L..47A}{\adsurllinklabel}

\bibitem[{{Aspin} {et~al.}(2006){Aspin}, {Barbieri}, {Boschi}, {Di Mille},
  {Rampazzi}, {Reipurth}, \& {Tsvetkov}}]{Aspin2006}
{Aspin}, C., {Barbieri}, C., {Boschi}, F., {Di Mille}, F., {Rampazzi}, F.,
  {Reipurth}, B., \& {Tsvetkov}, M. 2006, \aj, 132, 1298
 \href{http://adsabs.harvard.edu/abs/2006AJ....132.1298A}{\adsurllinklabel}

\bibitem[{{Aspin} {et~al.}(2008){Aspin}, {Beck}, \& {Reipurth}}]{Aspin2008}
{Aspin}, C., {Beck}, T.~L., \& {Reipurth}, B. 2008, \aj, 135, 423
 \href{http://adsabs.harvard.edu/abs/2008AJ....135..423A}{\adsurllinklabel}

\bibitem[{{Aspin} {et~al.}(2009){Aspin}, {Reipurth}, {Beck}, {Aldering},
  {Doering}, {Hammel}, {Lynch}, {Meixner}, {Pecontal}, {Russell}, {Sitko},
  {Thomas}, \& {U}}]{Aspin2009b}
{Aspin}, C., {Reipurth}, B., {Beck}, T.~L., {Aldering}, G., {Doering}, R.~L.,
  {Hammel}, H.~B., {Lynch}, D.~K., {Meixner}, M., {Pecontal}, E., {Russell},
  R.~W., {Sitko}, M.~L., {Thomas}, R.~C., \& {U}, V. 2009, \apjl, 692, L67


\bibitem[{{Audard} {et~al.}(2014){Audard}, {{\'A}brah{\'a}m}, {Dunham},
  {Green}, {Grosso}, {Hamaguchi}, {Kastner}, {K{\'o}sp{\'a}l}, {Lodato},
  {Romanova}, {Skinner}, {Vorobyov}, \& {Zhu}}]{Audard2014}
{Audard}, M., {{\'A}brah{\'a}m}, P., {Dunham}, M.~M., {Green}, J.~D., {Grosso},
  N., {Hamaguchi}, K., {Kastner}, J.~H., {K{\'o}sp{\'a}l}, {\'A}., {Lodato},
  G., {Romanova}, M.~M., {Skinner}, S.~L., {Vorobyov}, E.~I., \& {Zhu}, Z.
  2014, Protostars and Planets VI, 387
 \href{http://adsabs.harvard.edu/abs/2014prpl.conf..387A}{\adsurllinklabel}

\bibitem[{{Bally} \& {Reipurth}(2001)}]{Bally2001}
{Bally}, J. \& {Reipurth}, B. 2001, \apj, 546, 299
 \href{http://adsabs.harvard.edu/abs/2001ApJ...546..299B}{\adsurllinklabel}

\bibitem[{{Beckwith} {et~al.}(1990){Beckwith}, {Sargent}, {Chini}, \&
  {Guesten}}]{Beckwith1990}
{Beckwith}, S.~V.~W., {Sargent}, A.~I., {Chini}, R.~S., \& {Guesten}, R. 1990,
  \aj, 99, 924
 \href{http://adsabs.harvard.edu/abs/1990AJ.....99..924B}{\adsurllinklabel}

\bibitem[{{Beltr{\'a}n} {et~al.}(2016){Beltr{\'a}n}, {Cesaroni}, {Moscadelli},
  {S{\'a}nchez-Monge}, {Hirota}, \& {Kumar}}]{Beltran2016}
{Beltr{\'a}n}, M.~T., {Cesaroni}, R., {Moscadelli}, L., {S{\'a}nchez-Monge},
  {\'A}., {Hirota}, T., \& {Kumar}, M.~S.~N. 2016, \aap, 593, A49
 \href{http://adsabs.harvard.edu/abs/2016A%26A...593A..49B}{\adsurllinklabel}

\bibitem[{{Bjerkeli} {et~al.}(2016){Bjerkeli}, {van der Wiel}, {Harsono},
  {Ramsey}, \& {J{\o}rgensen}}]{Bjerkeli2016b}
{Bjerkeli}, P., {van der Wiel}, M.~H.~D., {Harsono}, D., {Ramsey}, J.~P., \&
  {J{\o}rgensen}, J.~K. 2016, \nat, 540, 406
 \href{http://adsabs.harvard.edu/abs/2016Natur.540..406B}{\adsurllinklabel}

\bibitem[{{Brice{\~n}o} {et~al.}(2004){Brice{\~n}o}, {Vivas}, {Hern{\'a}ndez},
  {Calvet}, {Hartmann}, {Megeath}, {Berlind}, {Calkins}, \&
  {Hoyer}}]{Briceno2004}
{Brice{\~n}o}, C., {Vivas}, A.~K., {Hern{\'a}ndez}, J., {Calvet}, N.,
  {Hartmann}, L., {Megeath}, T., {Berlind}, P., {Calkins}, M., \& {Hoyer}, S.
  2004, \apjl, 606, L123
 \href{http://adsabs.harvard.edu/abs/2004ApJ...606L.123B}{\adsurllinklabel}

\bibitem[{{Cieza} {et~al.}(2017){Cieza}, {Casassus}, {Hales}, {Perez},
  {Principe}, {Ruiz-Rodriguez}, {Williams}, {Zhu}, \& {Zurlo}}]{Cieza2017}
{Cieza}, L., {Casassus}, S., {Hales}, A., {Perez}, S., {Principe}, D.,
  {Ruiz-Rodriguez}, D., {Williams}, J., {Zhu}, Z., \& {Zurlo}, A. 2017, \mnras
  , in prep.


\bibitem[{{Cieza} {et~al.}(2016){Cieza}, {Casassus}, {Tobin}, {Bos},
  {Williams}, {Perez}, {Zhu}, {Caceres}, {Canovas}, {Dunham}, {Hales},
  {Prieto}, {Principe}, {Schreiber}, {Ruiz-Rodriguez}, \& {Zurlo}}]{Cieza2016}
{Cieza}, L.~A., {Casassus}, S., {Tobin}, J., {Bos}, S.~P., {Williams}, J.~P.,
  {Perez}, S., {Zhu}, Z., {Caceres}, C., {Canovas}, H., {Dunham}, M.~M.,
  {Hales}, A., {Prieto}, J.~L., {Principe}, D.~A., {Schreiber}, M.~R.,
  {Ruiz-Rodriguez}, D., \& {Zurlo}, A. 2016, \nat, 535, 258
 \href{http://adsabs.harvard.edu/abs/2016Natur.535..258C}{\adsurllinklabel}

\bibitem[{{Clarke} {et~al.}(1990){Clarke}, {Lin}, \& {Pringle}}]{Clarke1990}
{Clarke}, C.~J., {Lin}, D.~N.~C., \& {Pringle}, J.~E. 1990, \mnras, 242, 439
 \href{http://adsabs.harvard.edu/abs/1990MNRAS.242..439C}{\adsurllinklabel}

\bibitem[{{Cunningham} {et~al.}(2009){Cunningham}, {Moeckel}, \&
  {Bally}}]{Cunningham2009b}
{Cunningham}, N.~J., {Moeckel}, N., \& {Bally}, J. 2009, \apj, 692, 943
 \href{http://adsabs.harvard.edu/abs/2009ApJ...692..943C}{\adsurllinklabel}

\bibitem[{{Dunham} {et~al.}(2014){Dunham}, {Stutz}, {Allen}, {Evans},
  {Fischer}, {Megeath}, {Myers}, {Offner}, {Poteet}, {Tobin}, \&
  {Vorobyov}}]{Dunham2014}
{Dunham}, M.~M., {Stutz}, A.~M., {Allen}, L.~E., {Evans}, II, N.~J., {Fischer},
  W.~J., {Megeath}, S.~T., {Myers}, P.~C., {Offner}, S.~S.~R., {Poteet}, C.~A.,
  {Tobin}, J.~J., \& {Vorobyov}, E.~I. 2014, ArXiv e-prints: 1401.1809
 \href{http://adsabs.harvard.edu/abs/2014arXiv1401.1809D}{\adsurllinklabel}

\bibitem[{{Evans} {et~al.}(1994){Evans}, {Balkum}, {Levreault}, {Hartmann}, \&
  {Kenyon}}]{Evans1994}
{Evans}, II, N.~J., {Balkum}, S., {Levreault}, R.~M., {Hartmann}, L., \&
  {Kenyon}, S. 1994, \apj, 424, 793
 \href{http://adsabs.harvard.edu/abs/1994ApJ...424..793E}{\adsurllinklabel}

\bibitem[{{Evans} {et~al.}(2009){Evans}, {Dunham}, {J{\o}rgensen}, {Enoch},
  {Mer{\'{\i}}n}, {van Dishoeck}, {Alcal{\'a}}, {Myers}, {Stapelfeldt},
  {Huard}, {Allen}, {Harvey}, {van Kempen}, {Blake}, {Koerner}, {Mundy},
  {Padgett}, \& {Sargent}}]{Evans2009}
{Evans}, II, N.~J., {Dunham}, M.~M., {J{\o}rgensen}, J.~K., {Enoch}, M.~L.,
  {Mer{\'{\i}}n}, B., {van Dishoeck}, E.~F., {Alcal{\'a}}, J.~M., {Myers},
  P.~C., {Stapelfeldt}, K.~R., {Huard}, T.~L., {Allen}, L.~E., {Harvey}, P.~M.,
  {van Kempen}, T., {Blake}, G.~A., {Koerner}, D.~W., {Mundy}, L.~G.,
  {Padgett}, D.~L., \& {Sargent}, A.~I. 2009, \apjs, 181, 321
 \href{http://adsabs.harvard.edu/abs/2009ApJS..181..321E}{\adsurllinklabel}

\bibitem[{{Frank} {et~al.}(2014){Frank}, {Ray}, {Cabrit}, {Hartigan}, {Arce},
  {Bacciotti}, {Bally}, {Benisty}, {Eisl{\"o}ffel}, {G{\"u}del}, {Lebedev},
  {Nisini}, \& {Raga}}]{Frank2014}
{Frank}, A., {Ray}, T.~P., {Cabrit}, S., {Hartigan}, P., {Arce}, H.~G.,
  {Bacciotti}, F., {Bally}, J., {Benisty}, M., {Eisl{\"o}ffel}, J.,
  {G{\"u}del}, M., {Lebedev}, S., {Nisini}, B., \& {Raga}, A. 2014, Protostars
  and Planets VI, 451
 \href{http://adsabs.harvard.edu/abs/2014prpl.conf..451F}{\adsurllinklabel}

\bibitem[{{Hales} {et~al.}(2015){Hales}, {Corder}, {Dent}, {Andrews}, {Eisner},
  \& {Cieza}}]{Hales2015}
{Hales}, A.~S., {Corder}, S.~A., {Dent}, W.~R.~D., {Andrews}, S.~M., {Eisner},
  J.~A., \& {Cieza}, L.~A. 2015, \apj, 812, 134
 \href{http://adsabs.harvard.edu/abs/2015ApJ...812..134H}{\adsurllinklabel}

\bibitem[{{Hamaguchi} {et~al.}(2012){Hamaguchi}, {Grosso}, {Kastner},
  {Weintraub}, {Richmond}, {Petre}, {Teets}, \& {Principe}}]{Hamaguchi2012}
{Hamaguchi}, K., {Grosso}, N., {Kastner}, J.~H., {Weintraub}, D.~A.,
  {Richmond}, M., {Petre}, R., {Teets}, W.~K., \& {Principe}, D. 2012, \apj,
  754, 32
 \href{http://adsabs.harvard.edu/abs/2012ApJ...754...32H}{\adsurllinklabel}

\bibitem[{{Herbig}(1966)}]{Herbig1966}
{Herbig}, G.~H. 1966, Vistas in Astronomy, 8, 109
 \href{http://adsabs.harvard.edu/abs/1966VA......8..109H}{\adsurllinklabel}

\bibitem[{{Herbig}(1977)}]{Herbig1977}
---. 1977, \apj, 217, 693
 \href{http://adsabs.harvard.edu/abs/1977ApJ...217..693H}{\adsurllinklabel}

\bibitem[{{Kastner} {et~al.}(2004){Kastner}, {Richmond}, {Grosso}, {Weintraub},
  {Simon}, {Frank}, {Hamaguchi}, {Ozawa}, \& {Henden}}]{Kastner2004b}
{Kastner}, J.~H., {Richmond}, M., {Grosso}, N., {Weintraub}, D.~A., {Simon},
  T., {Frank}, A., {Hamaguchi}, K., {Ozawa}, H., \& {Henden}, A. 2004, \nat,
  430, 429
 \href{http://adsabs.harvard.edu/abs/2004Natur.430..429K}{\adsurllinklabel}

\bibitem[{{Kennedy} \& {Kenyon}(2008)}]{Kennedy2008}
{Kennedy}, G.~M. \& {Kenyon}, S.~J. 2008, \apj, 673, 502
 \href{http://adsabs.harvard.edu/abs/2008ApJ...673..502K}{\adsurllinklabel}

\bibitem[{{Kenyon} {et~al.}(1990){Kenyon}, {Hartmann}, {Strom}, \&
  {Strom}}]{Kenyon1990}
{Kenyon}, S.~J., {Hartmann}, L.~W., {Strom}, K.~M., \& {Strom}, S.~E. 1990,
  \aj, 99, 869
 \href{http://adsabs.harvard.edu/abs/1990AJ.....99..869K}{\adsurllinklabel}

\bibitem[{{Kraus} {et~al.}(2006){Kraus}, {Balega}, {Elitzur}, {Hofmann},
  {Preibisch}, {Rosen}, {Schertl}, {Weigelt}, \& {Young}}]{Kraus2006}
{Kraus}, S., {Balega}, Y., {Elitzur}, M., {Hofmann}, K.-H., {Preibisch}, T.,
  {Rosen}, A., {Schertl}, D., {Weigelt}, G., \& {Young}, E.~T. 2006, \aap, 455,
  521
 \href{http://adsabs.harvard.edu/abs/2006A%26A...455..521K}{\adsurllinklabel}

\bibitem[{{Langer} \& {Penzias}(1993)}]{Langer1993}
{Langer}, W.~D. \& {Penzias}, A.~A. 1993, \apj, 408, 539
 \href{http://adsabs.harvard.edu/abs/1993ApJ...408..539L}{\adsurllinklabel}

\bibitem[{{Lodato} \& {Clarke}(2004)}]{Lodato2004}
{Lodato}, G. \& {Clarke}, C.~J. 2004, \mnras, 353, 841
 \href{http://adsabs.harvard.edu/abs/2004MNRAS.353..841L}{\adsurllinklabel}

\bibitem[{{Lubow} \& {Ogilvie}(2000)}]{Lubow2000}
{Lubow}, S.~H. \& {Ogilvie}, G.~I. 2000, \apj, 538, 326
 \href{http://adsabs.harvard.edu/abs/2000ApJ...538..326L}{\adsurllinklabel}

\bibitem[{{McMullin} {et~al.}(2007){McMullin}, {Waters}, {Schiebel}, {Young},
  \& {Golap}}]{McMullin2007}
{McMullin}, J.~P., {Waters}, B., {Schiebel}, D., {Young}, W., \& {Golap}, K.
  Astronomical Society of the Pacific Conference Series, Vol. 376, ,
  Astronomical Data Analysis Software and Systems XVI, ed. R.~A.
  {Shaw}F.~{Hill} \& D.~J. {Bell}, 127
 \href{http://adsabs.harvard.edu/abs/2007ASPC..376..127M}{\adsurllinklabel}

\bibitem[{{McNeil} {et~al.}(2004){McNeil}, {Reipurth}, \& {Meech}}]{McNeil2004}
{McNeil}, J.~W., {Reipurth}, B., \& {Meech}, K. 2004, \iaucirc, 8284
 \href{http://adsabs.harvard.edu/abs/2004IAUC.8284....1M}{\adsurllinklabel}

\bibitem[{{Menten} {et~al.}(2007){Menten}, {Reid}, {Forbrich}, \&
  {Brunthaler}}]{Menten2007}
{Menten}, K.~M., {Reid}, M.~J., {Forbrich}, J., \& {Brunthaler}, A. 2007, \aap,
  474, 515
 \href{http://adsabs.harvard.edu/abs/2007A%26A...474..515M}{\adsurllinklabel}

\bibitem[{{Muzerolle} {et~al.}(2005){Muzerolle}, {Megeath}, {Flaherty},
  {Gordon}, {Rieke}, {Young}, \& {Lada}}]{Muzerolle2005}
{Muzerolle}, J., {Megeath}, S.~T., {Flaherty}, K.~M., {Gordon}, K.~D., {Rieke},
  G.~H., {Young}, E.~T., \& {Lada}, C.~J. 2005, \apjl, 620, L107
 \href{http://adsabs.harvard.edu/abs/2005ApJ...620L.107M}{\adsurllinklabel}

\bibitem[{{Ninan} {et~al.}(2013){Ninan}, {Ojha}, {Bhatt}, {Ghosh}, {Mohan},
  {Mallick}, {Tamura}, \& {Henning}}]{Ninan2013}
{Ninan}, J.~P., {Ojha}, D.~K., {Bhatt}, B.~C., {Ghosh}, S.~K., {Mohan}, V.,
  {Mallick}, K.~K., {Tamura}, M., \& {Henning}, T. 2013, \apj, 778, 116
 \href{http://adsabs.harvard.edu/abs/2013ApJ...778..116N}{\adsurllinklabel}

\bibitem[{{Offner} {et~al.}(2016){Offner}, {Dunham}, {Lee}, {Arce}, \&
  {Fielding}}]{Offner2016}
{Offner}, S.~S.~R., {Dunham}, M.~M., {Lee}, K.~I., {Arce}, H.~G., \&
  {Fielding}, D.~B. 2016, \apjl, 827, L11
 \href{http://adsabs.harvard.edu/abs/2016ApJ...827L..11O}{\adsurllinklabel}

\bibitem[{{Okuzumi} {et~al.}(2012){Okuzumi}, {Tanaka}, {Kobayashi}, \&
  {Wada}}]{Okuzumi2012}
{Okuzumi}, S., {Tanaka}, H., {Kobayashi}, H., \& {Wada}, K. 2012, \apj, 752,
  106
 \href{http://adsabs.harvard.edu/abs/2012ApJ...752..106O}{\adsurllinklabel}

\bibitem[{{Perez} {et~al.}(2015){Perez}, {Casassus}, {M{\'e}nard}, {Roman},
  {van der Plas}, {Cieza}, {Pinte}, {Christiaens}, \& {Hales}}]{Perez2015}
{Perez}, S., {Casassus}, S., {M{\'e}nard}, F., {Roman}, P., {van der Plas}, G.,
  {Cieza}, L., {Pinte}, C., {Christiaens}, V., \& {Hales}, A.~S. 2015, \apj,
  798, 85
 \href{http://adsabs.harvard.edu/abs/2015ApJ...798...85P}{\adsurllinklabel}

\bibitem[{{Pollack} {et~al.}(1994){Pollack}, {Hollenbach}, {Beckwith},
  {Simonelli}, {Roush}, \& {Fong}}]{Pollack1994}
{Pollack}, J.~B., {Hollenbach}, D., {Beckwith}, S., {Simonelli}, D.~P.,
  {Roush}, T., \& {Fong}, W. 1994, \apj, 421, 615
 \href{http://adsabs.harvard.edu/abs/1994ApJ...421..615P}{\adsurllinklabel}

\bibitem[{{Principe} {et~al.}(2014){Principe}, {Kastner}, {Grosso},
  {Hamaguchi}, {Richmond}, {Teets}, \& {Weintraub}}]{Principe2014}
{Principe}, D.~A., {Kastner}, J.~H., {Grosso}, N., {Hamaguchi}, K., {Richmond},
  M., {Teets}, W.~K., \& {Weintraub}, D.~A. 2014, \apjs, 213, 4
 \href{http://adsabs.harvard.edu/abs/2014ApJS..213....4P}{\adsurllinklabel}

\bibitem[{{Reipurth} \& {Aspin}(2004)}]{Reipurth2004}
{Reipurth}, B. \& {Aspin}, C. 2004, \apjl, 606, L119
 \href{http://adsabs.harvard.edu/abs/2004ApJ...606L.119R}{\adsurllinklabel}

\bibitem[{{Rettig} {et~al.}(2005){Rettig}, {Brittain}, {Gibb}, {Simon}, \&
  {Kulesa}}]{Rettig2005}
{Rettig}, T.~W., {Brittain}, S.~D., {Gibb}, E.~L., {Simon}, T., \& {Kulesa}, C.
  2005, \apj, 626, 245
 \href{http://adsabs.harvard.edu/abs/2005ApJ...626..245R}{\adsurllinklabel}

\bibitem[{{Ru{\'{\i}}z-Rodr{\'{\i}}guez}
  {et~al.}(2017{\natexlab{a}}){Ru{\'{\i}}z-Rodr{\'{\i}}guez}, {Cieza},
  {Williams}, {Principe}, {Tobin}, {Zhu}, \& {Zurlo}}]{Ruiz-Rodriguez2017a}
{Ru{\'{\i}}z-Rodr{\'{\i}}guez}, D., {Cieza}, L.~A., {Williams}, J.~P.,
  {Principe}, D., {Tobin}, J.~J., {Zhu}, Z., \& {Zurlo}, A. 2017{\natexlab{a}},
  \mnras, 468, 3266
 \href{http://adsabs.harvard.edu/abs/2017MNRAS.468.3266R}{\adsurllinklabel}

\bibitem[{{Ru{\'{\i}}z-Rodr{\'{\i}}guez}
  {et~al.}(2017{\natexlab{b}}){Ru{\'{\i}}z-Rodr{\'{\i}}guez}, {Cieza},
  {Williams}, {Tobin}, {Hales}, {Zhu}, {Mu{\v z}i{\'c}}, {Principe}, {Canovas},
  {Zurlo}, {Casassus}, {Perez}, \& {Prieto}}]{Ruiz-Rodriguez2017b}
{Ru{\'{\i}}z-Rodr{\'{\i}}guez}, D., {Cieza}, L.~A., {Williams}, J.~P., {Tobin},
  J.~J., {Hales}, A., {Zhu}, Z., {Mu{\v z}i{\'c}}, K., {Principe}, D.,
  {Canovas}, H., {Zurlo}, A., {Casassus}, S., {Perez}, S., \& {Prieto}, J.~L.
  2017{\natexlab{b}}, \mnras, 466, 3519
 \href{http://adsabs.harvard.edu/abs/2017MNRAS.466.3519R}{\adsurllinklabel}

\bibitem[{{Schwartz}(1982)}]{Schwartz1982}
{Schwartz}, P.~R. 1982, \apj, 252, 589
 \href{http://adsabs.harvard.edu/abs/1982ApJ...252..589S}{\adsurllinklabel}

\bibitem[{{Shepherd} {et~al.}(2000){Shepherd}, {Yu}, {Bally}, \&
  {Testi}}]{Shepherd2000}
{Shepherd}, D.~S., {Yu}, K.~C., {Bally}, J., \& {Testi}, L. 2000, \apj, 535,
  833
 \href{http://adsabs.harvard.edu/abs/2000ApJ...535..833S}{\adsurllinklabel}

\bibitem[{{Su} {et~al.}(2007){Su}, {Liu}, {Chen}, {Zhang}, \&
  {Cesaroni}}]{Su2007}
{Su}, Y.-N., {Liu}, S.-Y., {Chen}, H.-R., {Zhang}, Q., \& {Cesaroni}, R. 2007,
  \apj, 671, 571
 \href{http://adsabs.harvard.edu/abs/2007ApJ...671..571S}{\adsurllinklabel}

\bibitem[{{Tazzari} {et~al.}(2017){Tazzari}, {Testi}, {Natta}, {Ansdell},
  {Carpenter}, {Guidi}, {Hogerheijde}, {Manara}, {Miotello}, {van der Marel},
  {van Dishoeck}, \& {Williams}}]{Tazzari2017}
{Tazzari}, M., {Testi}, L., {Natta}, A., {Ansdell}, M., {Carpenter}, J.,
  {Guidi}, G., {Hogerheijde}, M., {Manara}, C.~F., {Miotello}, A., {van der
  Marel}, N., {van Dishoeck}, E.~F., \& {Williams}, J.~P. 2017, ArXiv e-prints:
  1707.01499
 \href{http://adsabs.harvard.edu/abs/2017arXiv170701499T}{\adsurllinklabel}

\bibitem[{{Teets} {et~al.}(2011){Teets}, {Weintraub}, {Grosso}, {Principe},
  {Kastner}, {Hamaguchi}, \& {Richmond}}]{Teets2011}
{Teets}, W.~K., {Weintraub}, D.~A., {Grosso}, N., {Principe}, D., {Kastner},
  J.~H., {Hamaguchi}, K., \& {Richmond}, M. 2011, \apj, 741, 83
 \href{http://adsabs.harvard.edu/abs/2011ApJ...741...83T}{\adsurllinklabel}

\bibitem[{{Terquem} {et~al.}(1999){Terquem}, {Eisl{\"o}ffel}, {Papaloizou}, \&
  {Nelson}}]{Terquem1999}
{Terquem}, C., {Eisl{\"o}ffel}, J., {Papaloizou}, J.~C.~B., \& {Nelson}, R.~P.
  1999, \apjl, 512, L131
 \href{http://adsabs.harvard.edu/abs/1999ApJ...512L.131T}{\adsurllinklabel}

\bibitem[{{Vorobyov} \& {Basu}(2005)}]{Vorobyov2005}
{Vorobyov}, E.~I. \& {Basu}, S. 2005, \apjl, 633, L137
 \href{http://adsabs.harvard.edu/abs/2005ApJ...633L.137V}{\adsurllinklabel}

\bibitem[{{Whelan} {et~al.}(2010){Whelan}, {Dougados}, {Perrin}, {Bonnefoy},
  {Bains}, {Redman}, {Ray}, {Bouy}, {Benisty}, {Bouvier}, {Chauvin}, {Garcia},
  {Grankvin}, \& {Malbet}}]{Whelan2010}
{Whelan}, E.~T., {Dougados}, C., {Perrin}, M.~D., {Bonnefoy}, M., {Bains}, I.,
  {Redman}, M.~P., {Ray}, T.~P., {Bouy}, H., {Benisty}, M., {Bouvier}, J.,
  {Chauvin}, G., {Garcia}, P.~J.~V., {Grankvin}, K., \& {Malbet}, F. 2010,
  \apjl, 720, L119
 \href{http://adsabs.harvard.edu/abs/2010ApJ...720L.119W}{\adsurllinklabel}

\bibitem[{{Williams} \& {Cieza}(2011)}]{Williams2011}
{Williams}, J.~P. \& {Cieza}, L.~A. 2011, \araa, 49, 67
 \href{http://adsabs.harvard.edu/abs/2011ARA%26A..49...67W}{\adsurllinklabel}

\bibitem[{{Zurlo} {et~al.}(2017){Zurlo}, {Cieza}, {Williams}, {Canovas},
  {Perez}, {Hales}, {Mu{\v z}i{\'c}}, {Principe},
  {Ru{\'{\i}}z-Rodr{\'{\i}}guez}, {Tobin}, {Zhang}, {Zhu}, {Casassus}, \&
  {Prieto}}]{Zurlo2017}
{Zurlo}, A., {Cieza}, L.~A., {Williams}, J.~P., {Canovas}, H., {Perez}, S.,
  {Hales}, A., {Mu{\v z}i{\'c}}, K., {Principe}, D.~A.,
  {Ru{\'{\i}}z-Rodr{\'{\i}}guez}, D., {Tobin}, J., {Zhang}, Y., {Zhu}, Z.,
  {Casassus}, S., \& {Prieto}, J.~L. 2017, \mnras, 465, 834
 \href{http://adsabs.harvard.edu/abs/2017MNRAS.465..834Z}{\adsurllinklabel}

\end{thebibliography}




\appendix

\begin{figure*}
\centering
\includegraphics[scale=0.6]{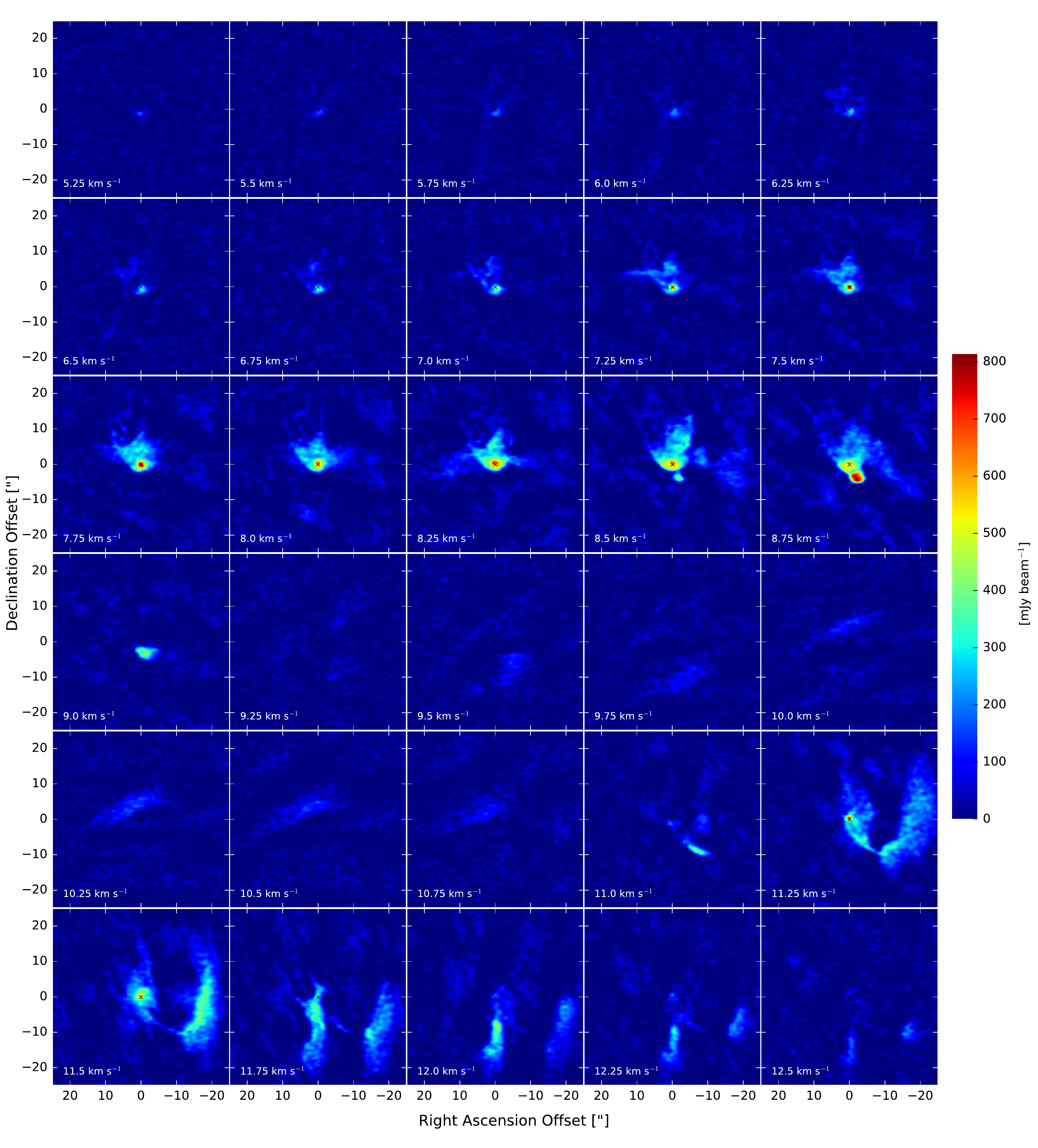}
\caption{ The complete $^{12}$CO channel map starting and ending at velocities (LSRK) where emission likely associated with V1647 Ori begins and ends, respectively.  The systemic velocity is $\sim$10 km s$^{-1}$. A black cross in each panel indicates the location of V1647 Ori and north is up and east is to the left.}
\label{}
\end{figure*}

\begin{figure*}
\centering
\includegraphics[scale=0.5]{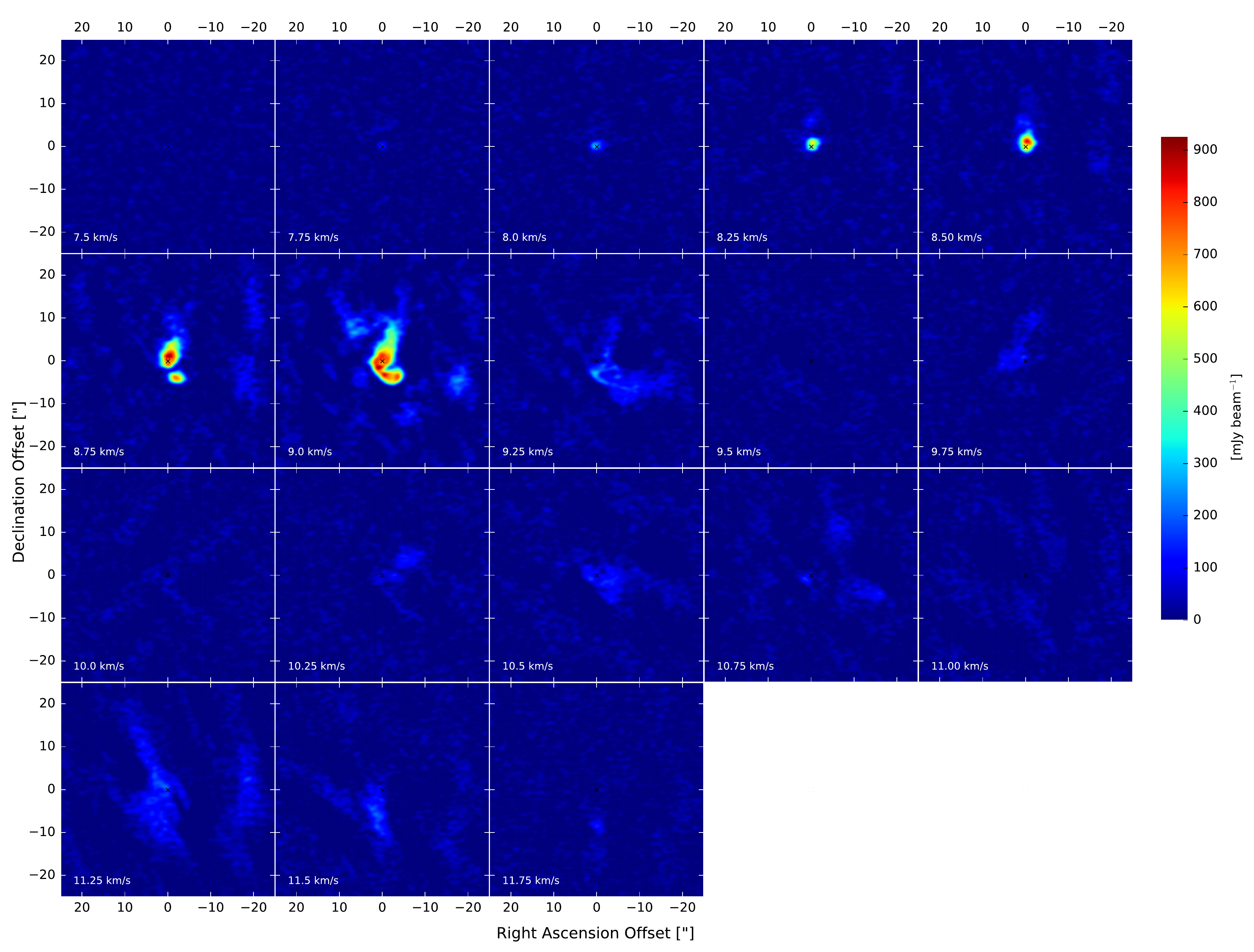}
\caption{ The complete $^{13}$CO channel map starting and ending at velocities (LSRK)  where emission likely associated with V1647 Ori begins and ends, respectively. The systemic velocity is $\sim$10 km s$^{-1}$. A black cross in each panel indicates the location of V1647 Ori and north is up and east is to the left.}
\label{}
\end{figure*}

\begin{figure*}
\centering
\includegraphics[scale=0.5]{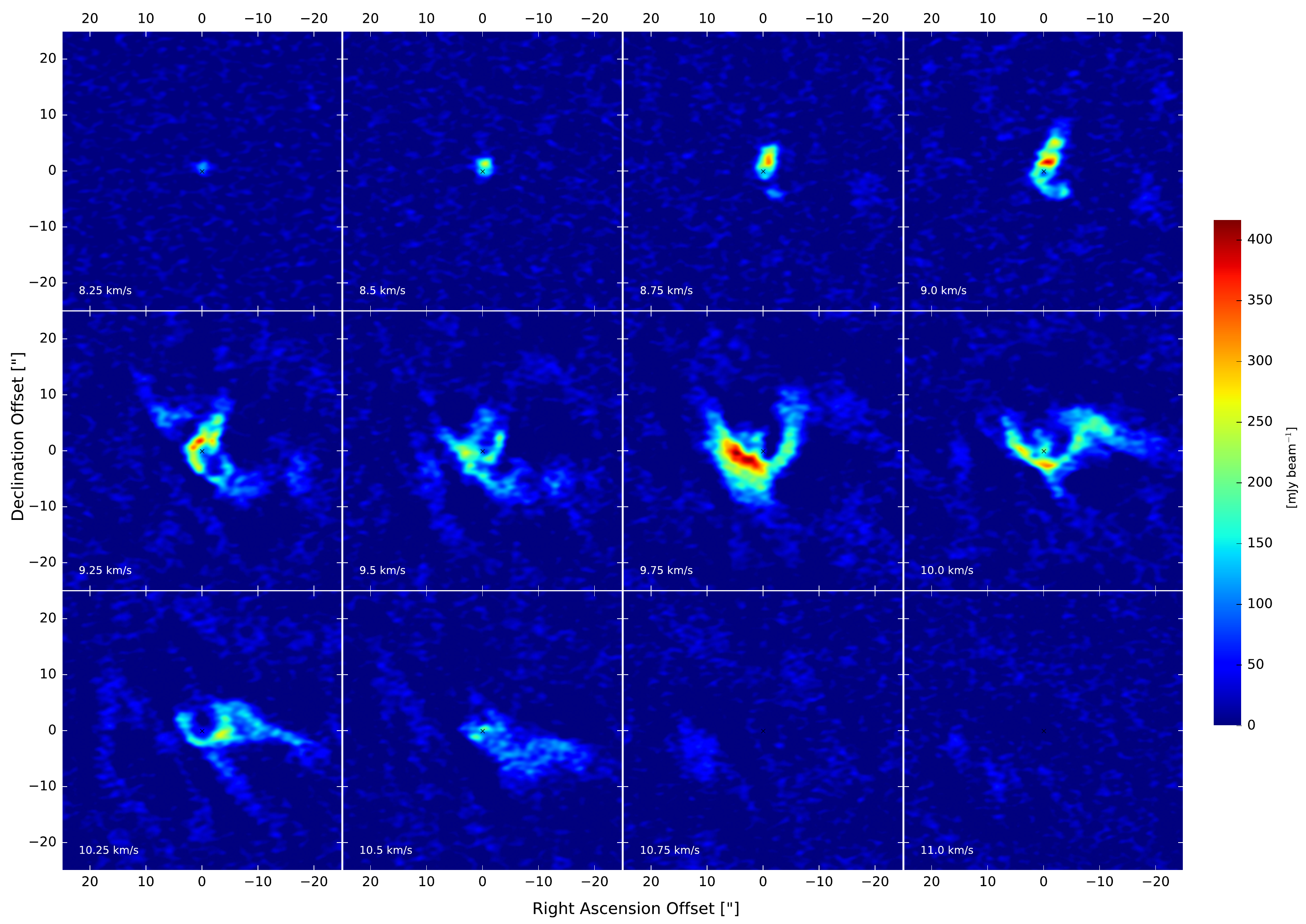}
\caption{The complete C$^{18}$O channel map starting and ending at velocities (LSRK)  where emission likely associated with V1647 Ori begins and ends, respectively. The systemic velocity is $\sim$10 km s$^{-1}$. A black cross in each panel indicates the location of V1647 Ori and north is up and east is to the left.}
\label{}
\end{figure*}




\label{lastpage}
\end{document}